\RequirePackage{rotating}

\documentclass[11pt,a4paper,fleqn]{article}
\usepackage[utf8]{inputenc}
\usepackage{authblk}
\usepackage{geometry}
\usepackage{xcolor}
\usepackage[round]{natbib}
\usepackage{pdflscape}  
\usepackage{appendix}
\usepackage{threeparttable}
\usepackage{booktabs}
\usepackage{subcaption}
\usepackage{multirow}
\usepackage{float}

\usepackage{amssymb}
\usepackage{amsbsy}
\usepackage{bm}
\usepackage{amsmath}
\usepackage{amsthm}
\usepackage{graphicx}
\usepackage{mathtools}
\usepackage{rotating}
\usepackage{setspace}
\usepackage{color, colortbl}
\definecolor{ashgrey}{rgb}{0.7, 0.75, 0.71}
\definecolor{columbiablue}{rgb}{0.61, 0.87, 1.0}
\definecolor{coral}{rgb}{1.0, 0.5, 0.31}

\definecolor{colBVAR}{HTML}{bababa}
\definecolor{colBART}{HTML}{d7191c}
\definecolor{colmixBART}{HTML}{fdae61}
\definecolor{colerrorBART}{HTML}{abd9e9}
\definecolor{colfullBART}{HTML}{2c7bb6}

\definecolor{colcons}{HTML}{e31a1c}
\definecolor{colSV}{HTML}{a6cee3}
\definecolor{colhBART}{HTML}{1f78b4}

\usepackage{enumitem}
\newlist{steps}{enumerate}{1}
\setlist[steps,1]{label = Step \arabic*:}

\usepackage{adjustbox}
\usepackage{dcolumn}
\newcolumntype{d}[1]{D..{#1}} 

\usepackage{caption}
\usepackage{subcaption}
\definecolor{nblue}{HTML}{000660}
\usepackage[colorlinks=true,urlcolor=nblue,linkcolor=nblue,citecolor=nblue]{hyperref}

\usepackage{cleveref}
\crefname{chapter}{Chapter}{Chapters}
\crefname{section}{Section}{Sections}
\crefname{subsection}{Section}{Sections}
\crefname{subsubsection}{Section}{Sections}
\crefname{figure}{Figure}{Figures}
\crefname{table}{Table}{Tables}
\crefname{equation}{Equation}{Equations}
\crefname{appendix}{Appendix}{Appendices}
\crefname{appendices}{Appendix}{Appendices}
\crefname{appsec}{Appendix}{Appendices}

\newcommand*{\myeqref}[2][Eq.~]{%
  \hyperref[{#2}]{#1(\ref*{#2})}%
}
\def\equationautorefname#1#2\null{%
  Eq.#1(#2\null)%
}

\begin{document}
\title{\textbf{Monetary policy and the joint distribution of income and wealth:\\
The heterogeneous case of the euro area
\large{}}}

\author{\large{
Anna \uppercase{Stelzer}}\thanks{
\noindent Anna Stelzer, Department of Economics, University of Salzburg. \textit{Address}: M\"{o}nchsberg 2, 5020 Salzburg, Austria. \textit{Email}: \href{mailto:anna.stelzer@plus.ac.at}{anna.stelzer@plus.ac.at}. I am thankful for helpful comments provided by Sylvia Frühwirth-Schnatter, Niko Hauzenberger, Florian Huber, Gary Koop, Michael Pfarrhofer and Michael Smith. I gratefully acknowledge funding by the Oesterreichische Nationalbank (project no. $18763$), the Austrian Science Fund (FWF, grant no. ZK $35$) and the financial support of the Humer Foundation. I am further acknowledging support by the Austrian Economics Association (NOeG) in the form of its dissertation fellowship.}
\\\vspace*{-0.5em}
\textit{University of Salzburg}}
\date{}

\maketitle\normalsize\vspace*{-3em}
\begin{center}

\vspace*{3em}\linespread{1.5}
\begin{minipage}{0.8\textwidth}
\noindent\small This papers aims to establish the empirical relationship between income, net wealth and their joint distribution in a selected group of euro area countries.
I estimate measures of dependence between income and net wealth using a semiparametric copula approach and calculate a bivariate Gini coefficient. By combining structural inference from vector autoregressions on the macroeconomic level with a simulation using microeconomic data, I investigate how conventional and unconventional monetary policy measures affect the joint distribution.
Results indicate that effects of monetary policy are highly heterogeneous across different countries, both in terms of the dependence of income and net wealth on each other, and in terms of inequality in both income and net wealth.

\end{minipage}
\end{center}

\clearpage\normalsize\doublespacing
\section{Introduction}\label{sec:intro}
Many economic theories treat income and wealth as alternative means to securing one's living, whereas an alternative perspective is that income and wealth are dependent on each other and that high/low levels of wealth imply high/low levels of income.
If income and wealth are in fact not simply alternative means of securing a standard of living but they are positively associated, then this would reinforce overall social inequality, and univariate measures of inequality for either income or wealth would only capture part of the story.
In fact, a greater positive association between income and wealth suggests that households are less able to rely on savings to smooth consumption.
In terms of economic policy, a positive association between income and wealth would imply that income taxation implicitly also targets wealth. 
Depending on political will and a household's position in the joint distribution of income and wealth, this may be desirable or not.
In any case, such relationships should be taken into account and care must be taken to avoid unintended side effects of fiscal or monetary policy.

Empirical evidence indeed suggests a positive association between income and wealth.
Using tax and financial accounts data for the US, \citet{saez2016wealth} show that the upsurge of top incomes is a key driver of the rapid increase in top wealth. 
Inequality in income and savings leads to inequality in wealth, and rising wealth inequality leads to higher capital income concentration.
This again yields bigger gaps between the top and the bottom of income and wealth.
\citet{JaenttiSierminskaSmeedingOECD2015} find a similar positive association between income and wealth for Canada, Germany, Italy, Sweden and the US for the years 1999, 2001 and 2002 and conclude that this positive association can only in part be explained by household characteristics like age or education.
Results by \citet{JaenntiSierminskaVanKerm2015} suggest that most cross-country differences for five OECD countries for 2007 and 2008 in a bivariate measure of inequality are not driven by differences in the dependence between income and wealth but are in fact determined by the underlying inequality in the wealth distribution.
\citet{fisher2022inequality} add another dimension to the analysis and conduct a share analysis of the distribution of income, consumption and wealth in the US between 1989 and 2016.
They find that inequality increased during that time in two and in three dimensions and that the increase in multidimensional inequality happened faster than in one-dimensional inequality.
\citet{Chauvel2018IncreasingII} study US data from 1995 to 2013 and examine the tails of the joint distribution of income and wealth.
Their results confirm previous findings that the link between income and wealth above the median has increased substantially over time.
Consequently, there is strong empirical evidence that income and wealth indeed depend positively on each other in many industrialized countries.
This suggests that a multidimensional perspective is crucial in order to properly assess inequality.

When it comes to the consequences that changes in monetary policy may imply for the distribution of both income and wealth, the literature is more comprehensive, but also more ambiguous.
\citet{colciago2019central} categorize the effects that monetary policy has on households into income, wealth and substitution effects.
As monetary policy influences interest rates, which savers receive and borrowers pay, and the value of both financial and real assets, it has a direct impact on households' income and their wealth.
This is captured in the income and wealth effect, respectively.
The substitution effect accounts for the fact that a change in the real interest rate affects the price of current versus future consumption.
The overall response of inequality to changes in monetary policy then depends on the relative importance of the three effects, how they interact with each other and heterogeneity among households.
A priori, it is therefore unclear what effect monetary policy has on the distribution of income and wealth.
The distribution of households along relevant dimensions such as wealth or income matters for the transmission of monetary policy on inequality, captured in what \citet{colciago2019central} call \textit{distributional channels}.
\citet{ampudia2018} group these channels in two broad categories, namely direct and indirect effects.
Direct effects stem from changes in households' incentive to save and their net financial income.
Indirect effects work through general equilibrium responses of both prices and wages after a monetary policy easing or tightening.\footnote{\citet{colciago2019central} provide a comprehensive literature review of both theoretical and empirical work on conventional and unconventional monetary policy as well as macroprudential policy.}

Early studies like \citet{romer1998} focused on the effect of inflation on inequality.
They conclude that sound monetary policy increases incomes of the poor in the long run by holding inflation low and keeping aggregate demand stable.
In the more recent literature, a distinction between effects of standard and unconventional monetary policy can be made.
For the case of conventional monetary policy, contractionary measures were found to increase income and earnings inequality in the US \citep{coibion2014}, the United Kingdom \citep[UK,][]{mumtaz2017}, the EA \citep[EA,][]{Guerello2018,samarina2019does} and a panel of advanced and emerging countries \citep{furceri2018}.
As interest rates increase during a monetary contraction, economic activity decreases which lowers the demand for labor and unemployment increases.
This in turn depresses wages which affects low-income households the most, as they rely heavily on labor income.
In contrast to those findings, \citet{inui2017} find that expansionary, and not contractionary, monetary policy increased inequality in Japan before the 2000s.
Some evidence, like \citet{ofarrell2016} or \citet{furceri2018}, also suggests that effects on monetary policy on income inequality may vary depending on the business cycle.
Another possible explanation for these partly contradictory results is provided by \citet{albert2020effects}, who find that opposing channels can counteract each other at times.
When it comes to the effects of conventional monetary policy on wealth inequality, most papers find that an expansionary policy reduces wealth inequality (\citet{doepke2006} for the US 1952--2004, \citet{ofarrell2016} for eight OECD countries 2007--2012 and \citet{inui2017} for Japan 1981--1998.
Empirical research on unconventional monetary policy like large-scale asset purchases finds that such programs might decrease income inequality by stimulating economic activity and wage growth in the EA \citep{Guerello2018, lenza2021}, Italy \citep{casiraghi2018} and the US \citep{bivens2015}. 
Different approaches that focus on the fact that unconventional monetary policy increases asset prices find that it might actually increase income inequality.
Higher asset prices tend to further increase capital income in the upper part of the distribution, therefore simultaneously increasing income inequality.
While evidence on the effects of monetary policy on income inequality is mixed, most studies find only negligible effects on wealth inequality \citep{casiraghi2018, lenza2021}.
\citet{alisdairwolf2023}, who focus on a third possible dimension of inequality, consumption, similarly conclude that inequality in consumption is hardly affected by changes in monetary policy, as different channels tend to balance each other out.

The aim of this paper is to contribute to the ongoing discussion on the effects of both conventional and unconventional monetary policy on the joint distribution of income and wealth in the EA, both in terms of inequality and in terms of the association between income and wealth.
First, it not only provides insights about the effect of monetary policy on the marginal distributions of both income and wealth, but more importantly, their joint distribution.
As previously explained, a focus solely on income or wealth inequality individually tends to underestimate overall inequality.
While there is a considerable amount of evidence on the relationship between monetary policy and either income or wealth, to my knowledge, there is no analysis that considers their joint distribution.
Second, this paper contributes to the existing literature by exploring dynamics and interdependencies between income and wealth in the EA, as the majority of existing papers focuses on the US or other non-European countries.\footnote{\citet{lenza2021} is a notable exception for this, although this paper analyses income and wealth separately.}
With the modeling approach chosen for this paper, I am able to account for heterogeneity in both EA-wide monetary policy effects and for cross-country differences in its transmission mechanism.

From a methodological perspective, by employing a semiparametric Bayesian copula approach, some shortcomings of traditional maximum likelihood estimation can be overcome.
As a consequence I am able to establish whether monetary policy effects on income and wealth either reinforce each other or work in opposite directions.

The remainder of this paper is structured as follows. 
Section \ref{sec:econometrics} first provides an overview of the empirical strategy of this paper before going into details of its individual steps.
Section \ref{sec:results} contains results of each step of the empirical strategy, a descriptive characterization of the joint distribution of income and net wealth, effects of monetary policy shocks on a macroeconomic level and their mapping into household data as well as the resulting effects on the joint distribution of income and net wealth.
Section \ref{sec:conclusion} concludes the paper.

\section{Empirical framework}\label{sec:econometrics}
\begin{figure}[h]
    \centering
    \includegraphics[width=\textwidth, trim=30 55 30 55, clip]{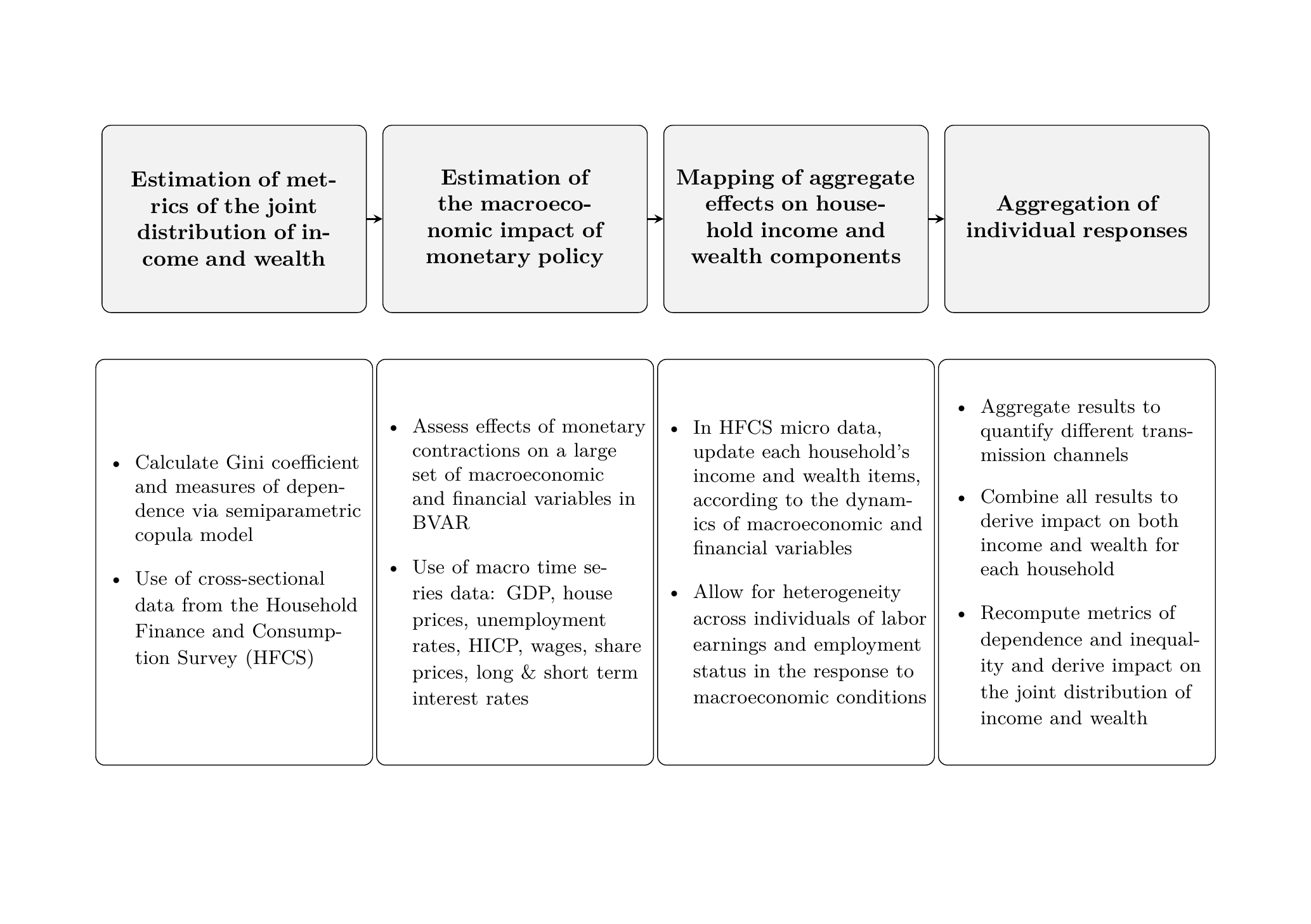}
    \caption{Overview of the four steps of the empirical strategy, data sources and level of analysis.}
    \label{fig:workflow}
\end{figure}

This section elaborates on the empirical strategies that this paper employs. 
It first explores the general strategy and data before going into detail on the individual steps as illustrated in Figure \ref{fig:workflow}, as well as provides details on the individual steps taken.

\subsection{Overview of the empirical strategy and data} \label{sec:overview}
The first step of the analysis is to establish the empirical relationship between income, net wealth and their joint distribution by estimating measures of dependence as well as calculating bivariate metrics of inequality. 
Using microdata from the second wave of the Household Finance and Consumption Survey \citep[HFCS,][]{HFCS} for Austria, Belgium, Germany, Spain, France, Germany, Greece, Ireland and Italy,\footnote{The second wave of the HFCS covers 18 EA countries, as well as Poland and Hungary. The selected sample of countries in this paper covers EA countries that are subject to the ECB's policy. The listed countries are chosen for illustrative purposes and are selected with the goal of creating a balanced sample of both core and periphery countries of different sizes.} covering the period between 2013 and the first half of 2015, I am able to establish empirical links between income and wealth on a household level. 

Besides providing descriptive measures such as a bivariate Gini coefficient, as proposed in \citet{grothe2022multivariate}, and illustrations of the joint distribution of income and wealth, the goal of this paper is to make inference on the dependence between income and net wealth.
I employ a Bayesian semiparametric copula model proposed by \citet{grazian2017} in order to estimate dependence metrics, namely Spearman's $\rho$, which characterizes overall dependence, and the tail dependence measures $\lambda_U$ and $\lambda_L$, for the upper and lower tail respectively.

This approach has several advantages compared to previous papers.
First, copulas are a flexible tool for modeling complex relationships between variables in a simple way. 
A copula approach allows to separate the problem of estimating the marginal distributions and the joint distribution as follows.
In a first step, one can estimate an appropriate distributional model for the marginal distributions of both income and net wealth.
In a second step, one uses these estimates to make inference on the parameters of a pre-specified copula function \citep[see][for more details]{joe2014}.
While copula functions have been previously used to model the joint distribution of income and wealth \citep[see, for example,][]{JaenntiSierminskaVanKerm2015}, using a semiparametric approach instead of choosing a specific copula function brings one decisive advantage.
For the joint distribution of income and net wealth, there is no clear theoretical implication for a specific functional form.
Indeed, most copula functions used in the literature seem to fall short in producing an acceptable model for the empirical data on income and wealth, as discussed in \citet{JaenttiSierminskaSmeedingOECD2015}.
By using a semiparametric copula approach, the choice of a copula function can be circumvented while still being able to estimate metrics depending on the (unspecified) copula.

Building on these results, I connect the aggregate effects of monetary policy to the joint distribution of income and wealth.
Microdata again stems from the HFCS, covering the EA for the years 2013--2015 \citep{HFCS}.
This data source is particularly useful, as it provides detailed information on different components of households' income and wealth, which allows to distinguish between heterogeneous developments between, e.g., labor and financial income or real and financial assets.
By making these distinctions, it is possible to get a clearer picture on which distributional channels mentioned in the introduction dominate and their empirical magnitudes.
Then, using an approach similar to \citet{casiraghi2018} and \citet{lenza2021}, the effects of monetary policy on income and wealth are evaluated in three steps.

First, the effects of monetary policy on the aggregate components of household income and wealth, both real and financial, are estimated in a Bayesian vector autoregressive model (BVAR). 
Using impulse response analysis, I investigate both short and long term effects of monetary policy on key series such as wages, the unemployment rate, interest rates and stock and house prices.
Second, these aggregate effects are mapped into microdata and evaluated on the household level.
This is achieved by distributing them on components of income and wealth estimated in the VAR according to the portfolios and characteristics of each household.
Third, all effects on income and wealth on an individual level are aggregated to describe the overall changes in the joint distribution of income and net wealth and inequality measured in terms of both variables.
In order to do so, I re-estimate the correlation measure between income and net wealth based on the copula model described in detail below, as well as a bivariate measure of inequality.\footnote{As discussed in more detail below, tail dependence measures are notorious for high estimation uncertainty due to the fact that there only few observations available in the tails. It is therefore not meaningful to estimate them after adding further uncertainty in the simulation process and, as a consequence, they are omitted from this part of the analysis.}
In the last step, I compare the metrics from before and after monetary policy shocks and evaluate the effects in different countries of the EA.

\subsection{Copula models and Empirical Likelihood} \label{sec:copula}
As mentioned above, copula models are flexible tools for modeling complex relationships between variables in a simple way by linking two or more univariate distributions via the copula function, resulting in their joint distribution.
Let $X_1$ and $X_2$ be two random variables, in this application income and net wealth, with marginal distribution functions $F_1(x_1; \tau_1)$ and $F_2(x_2; \tau_2)$.
A copula is an alternative representation of the joint distribution of $X=(X_1, X_2)$. \citet{sklar1959fonctions} shows that there always exists a bivariate function $C_{\theta}:[0,1]^2 \rightarrow [0,1]$ s.t.
\begin{equation*}
    F(x_1, x_2; \tau, \theta)=C_{\theta}(F_1(x_1;\tau_1), F_2(x_2;\tau_2)),
\end{equation*}
where $\tau = (\tau_1, \tau_2)$ is a vector of parameters of the marginal distributions and $\theta$ is a parameter governing the joint distribution $F$. $C_{\theta}$ is itself a distribution function with uniform margins on $[0,1]$.\footnote{Note that the copula function $C_{\theta}$ does not determine the marginal distribution of $F$, but accounts for dependence between $X_1$ and $X_2$.}
In case both marginals $F_1(x_1; \tau_1)$ and $F_2(x_2; \tau_2)$ are continuous, the density of $(X_1, X_2)$ has the unique copula representation
\begin{equation*}
    f(x|\tau, \theta)=c(u;\theta) f_1(x_1|\tau_1) f_2(x_2|\tau_2),
\end{equation*}
with $u=(u_1, u_2)=(F_1(x_1;\tau_1), F_2(x_2;\tau_2))$ and $c(u; \theta)$ being the derivative of $C_{\theta}$.
Most popular (frequentist) methods rely on a two-stage maximum likelihood  estimation in which parameters of the marginal distribution are estimated first, resulting in point estimates of $\tau$, which are in turn used to generate pseudo-data \citep{joe2005}.
Inference on the copula parameters is then based on the pseudo-data. 
Within this two-step approach, it is not possible to account for uncertainty in the estimation of the marginal parameters. 
Bayesian methods can alleviate this problem by considering posterior distributions instead of point estimates.
Given a prior $\pi(\theta, \tau_1, \tau_2)$ and a sample of $n$ independent observations $(x_{i1}, x_{i2})$ for $i=1, \ldots, n$ (in this application $n$ is given by the number of households in a country's HFCS sample), the posterior distribution for the parameter vector is
\begin{equation*}
    \pi(\theta, \tau | x) \propto \pi(\theta,\tau) \prod_{i=1}^{n} \big[c(u_i;\theta) f_1(x_{i1}|\tau_1) f_2(x_{i2}|\tau_2) \big].
\end{equation*}
The two step procedure of \citet{joe2005} can then easily be modified in a Bayesian framework by evaluating $\pi(\theta, \tau_1, \tau_2|x)$ with the help of a Markov chain Monte Carlo (MCMC) algorithm, with $\theta$ and $(\tau_1, \tau_2)$ generated separately.
When the goal of analysis is not the complete dependence structure but instead functions of the copula function, it is possible to follow the semiparametric approach proposed in \cite{grazian2017}.
In this approach, the marginal distributions $F_1$ and $F_2$ follow a parametric model, yet no parametric assumptions for the copula function $C_\theta$ have to be made.
Instead, a given function $\psi$ of interest of $C_\theta$, $\psi(C_\theta)$ is estimated by approximating the posterior distribution of $\psi$ by
\begin{equation*}
    \pi(\psi|x) \propto \pi(\psi)\hat{L}(\psi;x),
\end{equation*}
where $\hat{L}(\psi;x)$ is a nonparametric approximation of the likelihood function for $\psi$, and $\pi(\psi)$ is the prior distribution. 
Specifically, $\hat{L}(\psi;x)$ is given by the Bayesian Exponentially Tilted Empirical Likelihood as in \citet{Schennach2005} and is estimated in the algorithm proposed by \citet{grazian2017}.
The copula function itself is of no particular interest for the purposes of this paper. 
Instead, interest lies in characteristics of the joint distribution, hence, functions of the copula function, such as Spearman's $\rho$, $\lambda_U$ and $\lambda_L$, which makes this semiparametric approach a suitable method for this application.

As for the parametric marginal distribution functions $F_1$ and $F_2$, there are several possible candidate models for income and net wealth, respectively.
I consider two candidate distributions for each $F_1$, income, and $F_2$, net wealth.
The literature on modeling income distributions is vast, and the Singh-Maddala distribution \citep{singh1976}, and the Dagum distribution \citep{dagum2008inc} are two of its champions.
The Singh-Maddala distribution is given by
\begin{equation*}
    F_1(x_1)=1-\left[1+\left( \frac{x_1}{b} \right)^a\right]^{-q},\label{model:sinmad}
\end{equation*}
and the related Dagum distribution is specified as
\begin{equation*}
    F_2(x_1)=\left[1+\left( \frac{x_1}{c} \right)^{-d}\right]^{-p}.\label{model:dagum}
\end{equation*}
Abstracting from the fact that financial income can become negative,\footnote{Occurences of negative income in the HFCS data is generally sparse, with five households in Germany (0.11 percent of the total sample), 73 households in Spain (1.20 percent), 95 households in France (0.79 percent), 35 households in Ireland (0.65 percent) and none in the remaining countries.} both distributions have support on $0 < x_1 < \infty$.
Finding suitable distributions for modeling net wealth is less straight forward. 
I consider the shifted log-normal distribution, also called three-parameter log-normal, which is given by
\begin{equation*}
    F_3(x_2)=\Phi \left(\frac{ln(x_2-\gamma)-\mu}{\sigma}\right).\label{model:lognormal3}
\end{equation*}
The last candidate distribution for net wealth is an adapted version of the $\kappa$-generalized distribution proposed by \citet{clementi2012generalized}, a mixture distribution which consists of a Weibull distribution for the negative values of net wealth and a Dagum distribution for its positive values:
\begin{align*}
    &F_{4}(x_2)= \omega_1 F_{3,1}(x_2) + \omega_2 F_{3,2}(x_2) + \omega_3 F_{3,3}(x_2), \quad \omega_1+\omega_2=\kappa, \quad \omega_3=1-\kappa,\\
    &F_{4,1}(x_2)= \omega_1 \left(1-e^{-(\frac{|x_2|}{l})^k} \right), \quad x_2 < 0\\
    &F_{4,2}(x_2)= \kappa, \quad x_2=0,\\
    &F_{4,3}(x_2)= \kappa + (1-\kappa) \left[1+\left( \frac{x_1}{c} \right)^{-d}\right]^{-p}, \quad x_2>0.\label{model:D3}
\end{align*}
The parameters of these marginal distributions are estimated using a random walk Metropolis–Hastings algorithm using independent priors. 
For non-negative parameters, $a,b,c,d,p,q,\sigma,l,k> 0$, $x_2 > \gamma \geq 0$, I use weakly informative inverse Gamma priors, and Gaussian priors for parameters with domain $\mathbb{R}$, $-\infty < \mu < \infty$.
Model fit of the candidate models is evaluated and compared using the Bayesian information criterion (BIC) and the deviance information criterion (DIC).\footnote{Note that both theoretically the approach proposed by \citet{grazian2017} allows to estimate the marginal distributions either parametrically or nonparametrically. In their paper, however, the authors concluded that as long as the choice of marginal distributions is sensible it does not affect the posterior distribution of $\psi(C_\theta)$, therefore a simple parametric estimation of the marginal distributions is chosen in this paper.}

\subsection{The macroeconomic impact of monetary policy shocks} \label{sec:macroMP}
To estimate aggregate effects of monetary policy shocks, I estimate country-specific BVAR models.
Let $\bm{y}_{t} = (y_{1t},\hdots,y_{Mt})'$ denote an $M\times1$-vector of endogenous variables, and define $\bm{z}_{t} = (\bm{y}_{t-1}',\hdots,\bm{y}_{t-P}')'$ as a stacked $K\times1$-vector ($K=MP$) comprised of $P$ lags of the endogenous variables. 
The model is then given by
\begin{equation*}
    \bm{y}_{t} = \bm{A}\bm{z}_{t} + \bm{\epsilon}_{t},\quad \bm{\epsilon}_{t}\sim\mathcal{N}(\bm{0}, \bm{\Sigma}) \label{model:linear}
\end{equation*}
where $\bm{A}$ is an $M\times K$-matrix of VAR coefficients, and $\bm{\epsilon}_t$ is a Gaussian error term with an $M\times M$ covariance matrix $\bm{\Sigma}$.
The relevant endogenous variables include country-specific unemployment rates, real GDP, house prices, an inflation measure and wages, as well as short- and long-term interest rates, spreads and stock prices for the EA.
All macroeconomic data is on a quarterly level and retrieved from the OECD, the ECB's statistical data warehouse (SWD) and Eurostat.\footnote{Table \ref{tab:dataVAR}, providing details on data definitions, sources and transformations, can be found in the appendix.}

The paper differentiates between standard monetary policy shocks and quantitative easing (QE) measures for each country in the sample.
For identification, I rely on zero-restrictions on contemporaneous responses, which can be achieved by a Cholesky decomposition of $\bm{\Sigma}$.
For standard monetary policy effects, I consider a shock to the short term interest rate, while I interpret a shock to the EA spreads as an EA-wide QE shock.\footnote{EA spreads are calculated as the difference between each country's and the German long term interest rate (usually perceived as risk-free) and applying a principal component (PC) analysis. The first PC is then considered to be the aggregate EA spread.}
This identification strategy follows the idea that large scale purchasing programs of longer-term securities aim to compress the yield curve in the EA, as proposed in \citet{baumeister2013unconventional, feldkircher2020} and \citet{WALENTIN2014}.
To complete the Bayesian set-up, I use a standard Horseshoe prior setup \citep{carvalho2010horseshoe} in the estimation framework to impose data-driven shrinkage and obtain meaningful and precise structural inference.

\subsection{Mapping of aggregate effects on household income and wealth components} \label{sec:microsim}
Aggregate responses estimated as in Section \ref{sec:macroMP} are used to simulate which effects monetary policy shocks have on households' portfolios and incomes.
However, as the goal is to differentiate both direct and indirect channels of monetary policy, several steps have to be taken.
First, using the information about how monetary policy shocks affect aggregate unemployment, the microdata is adjusted for that change, both in terms of unemployment status and increased income.
Second, direct effects of monetary policy changes on wages and wealth assets have to be accounted for.

\begin{table}[ht]
    \centering
    \caption{Strategy for adjusting income and wealth components in the macro-based microsimulation.}
    \label{tab:simtab}
    \begin{tabular}{ll}
    \toprule
    \multicolumn{2}{c}{Net wealth assets}\\
    \hline
        Main residence $x_{i2,1}$ & $x_{i2,1} \times \Delta(\text{house prices})_{h}$ \\
        Other real estate $x_{i2,2}$ & $x_{i2,2} \times \Delta(\text{house prices})_{h}$\\
        Self-employment business $x_{i2,3}$ & no adjustment\\
        Publicly traded shares $x_{i2,4}$ & $x_{i2,4} \times \Delta(\text{stock prices})_{h}$\\
        Bonds $x_{i2,5}$ & $x_{i2,5} \times \Delta(\text{bond prices})_{h}$\\
        Voluntary pension or whole life insurance $x_{i2,6}$ & no adjustment\\
        Deposits $x_{i2,7}$ & no adjustment\\
        Other financial assets $x_{i2,8}$ & no adjustment\\
        Total liabilities $x_{i2,9}$ & no adjustment\\
    \bottomrule
     \multicolumn{2}{c}{Income component}\\
     \hline
        Employment income $x_{i1,1}$ & if person becomes employed: imputation\\
                                     & otherwise: $x_{i1,1} \times \Delta(\text{wages})_{h}$\\
        Self-employment income $x_{i1,2}$ & $x_{i1,2} \times \Delta(\text{wages})_{h}$\\
        Income from pensions $x_{i1,3}$ & no adjustment\\
        Rental income from real estate property $x_{i1,4}$ & no adjustment\\
        Income from financial investments $x_{i1,5}$ & no adjustment\\
        Unemployment benefits and transfers $x_{i1,6}$ & set to zero if employment income is imputed,\\
                                                        & otherwise no adjustment\\
    \bottomrule
    \end{tabular}
\end{table}

\subsubsection{The indirect channel of monetary policy}
The indirect channel refers to the effect that a monetary policy shock affects the general equilibrium and can result in changes in employment via changes in wages and prices.
The VAR from Section \ref{sec:macroMP} allows to derive the number of people who find or lose employment after a monetary policy shock in a given economy from the IRFs of unemployment ($\delta_h=\Delta(unemployment)_{h}$) for horizon $h=1,\hdots,12$.
This information can in turn be used for adjusting the number of unemployed people in the HFCS sample.

This procedure consists of two steps. 
First, I estimate the probability of being employed for every person in the sample and adjust the labor status in accordance to those probabilities as well as the overall number of people who lose or find employment after a monetary policy shock.
Second, income of all people (and as a consequence, households) who experienced a change in their labor status is adjusted.
If employment is lost, unemployment benefits replace employment income.
If a person's labor status switches from unemployed to employed, an imputed employment income replaces unemployment benefits.

For estimating the probability of being employed for each person in my sample I rely on a probit Bayesian additive regression tree (pBART) model, a nonparametric machine learning method developed by \citet{chipman2010}.
Nonparametric models like BART are useful new statistical tools which are able to successfully model non-linear relationships and \citet{chipman2010} find that BART is able to make very accurate predictions in a wide variety of datasets and applications.

For an individual $j$ from household $i$, let $z_{i}^j$ be their labor status, which is zero when the person is unemployed and one when he or she is employed.
Further, let $\bm{x}_{i}^j$ be a vector containing demographic characteristics which influence the labor status of a person, namely gender, the highest level of education attained, age, marital status and the number of dependent children.
Then, pBART is defined as
\begin{equation*}
    p(\bm{x}_{i}^j) \equiv P[z_{i}^j=1|\bm{x}_{i}^j] = \Phi[G(\bm{x}_{i}^j)], \label{model:pBART}
\end{equation*}
where $\Phi[\cdot]$ is the standard normal c.d.f. and each classification probability $p(\bm{x}_{i}^j)$ is given by a function of $G(\bm{x}_{i}^j)$, a sum of regression trees such that
\begin{equation*}
    G(\bm{x}_{i}^j) \equiv \sum_{m=1}^M g(\bm{x}_{i}^j; T_m, K_m). \label{model:pBART_trees}
\end{equation*}
$T_m$ denotes a binary decision tree which consists of interior node decision rules and a set of terminal nodes, and $K_m$ is the set of parameter values associated with each of the terminal nodes of $T_m$.
The splitting rules associated with each tree $T_m$ are binary splits based on a selected characteristic and a threshold chosen by the model.

If unemployment decreases at a given IRF horizon $h$, the $\delta_h$ persons with the greatest probability of being employed which are, however, unemployed, now get assigned the labor status employed, implying that they found a position as a consequence of the monetary policy shock.
This means of course that their income switches from unemployment benefits to labor income, which is assumed to be higher.
In order to account for this fact, I again employ BART to impute income from employment for this group.
The reason for choosing BART over more conventional imputation methods once more is that BART is a highly flexible tool which relies on very little assumptions. 
Additionally, as interest lies in predicting income levels and not making inference about parameters of a wage equation, choosing a nonparametric method fits well into the framework of other non- or semiparametric methods used in this paper.
For imputing employment income $x_{i1,1}^j$ for person $j$ in household $i$ in the microsimulation, BART takes the following form:
\begin{equation*}
   x_{i1,1}^j = \sum_{m=1}^M g(\bm{x}_{i}^j; T_m, K_m)+\epsilon_i^j, \quad \epsilon_i^j \sim N(0,\sigma^2),
   \label{model:BART}
\end{equation*}
where $\bm{x}_{i}^j$ again is a vector of relevant predictors of a person's (potential) employment income, i.e., gender, highest level of education attained, age, marital status, number of dependent children and time that a person was employed at his or her current job as a proxy for experience.
$\epsilon_i^j$ is a standard Gaussian error term and $T_m$ and $K_m$ again denote the terminal nodes of given tree and the parameter values that are associated with each terminal node, respectively.

Conversely, persons with a low probability of being employed but who are nevertheless employed lose their employed status and are instead assigned unemployment if the macroeconomonic results imply that aggregate unemployment increased at a given time horizon.
The modeling assumption for this group is that they lose their employment income and instead receive unemployment benefits.
These are calculated based on average unemployment benefit gross replacement rates \citep{oecd_unemp2021}.

Note that while the employment effect of monetary policy is simulated on an individual level, I aggregate these effects back to a household level, thus resulting in changes in the distribution of household income.

\subsubsection{The direct channel of monetary policy}
After gauging the response of corresponding aggregate measures from the impulse response analysis, I multiply the respective income components and wealth assets with the estimated changes from the IRFs, which constitutes the direct effects of monetary policy due to changes in incentives and financial income for the households.
Table \ref{tab:simtab} gives an overview of the relevant income and net wealth components and on how aggregate responses are distributed into the microdata.
Let income of a household $i$ be the sum of all six income components, $x_{i1} = \sum_{k=1}^6 x_{i1,k}$, and household $i$'s net wealth the sum of all eight possible assets as well as the total debt of a household $x_{i2} = \sum_{k=1}^9 x_{i2,k}$, for $i=1,\hdots,n$ households of the HFCS sample.
If applicable, the components of income (employment and self-employment income) and the asset classes of net wealth (main residence and other real estate property, shares and bonds\footnote{Bonds are multiplied with the change in stock prices as implied by the change in the long term interest rate.}) are multiplied with respective IRF change at each time horizon.\footnote{This approach assumes that households do not rebalance their portfolios after a monetary policy shock. This is, of course, a simplifying assumption, but some degree of inertia is supported by previous findings in the literature \citep[see, e.g.,][]{ameriks2004household,brunnermeier2008wealth}.}
After re-aggregating both household income and net wealth for each horizon $h$, I then recompute the measures of dependence and inequality from Section \ref{sec:copula} and compare them to their initial values to draw conclusions about the effect of monetary policy on the joint distribution of income and wealth.

\section{Empirical results}
\label{sec:results}

\subsection{Bivariate inequality and the distribution of income and net wealth}
\label{sec:descresults}

\begin{sidewaysfigure}
    \begin{minipage}[t]{.5\textwidth}
      \centering
      \includegraphics[scale=0.5]{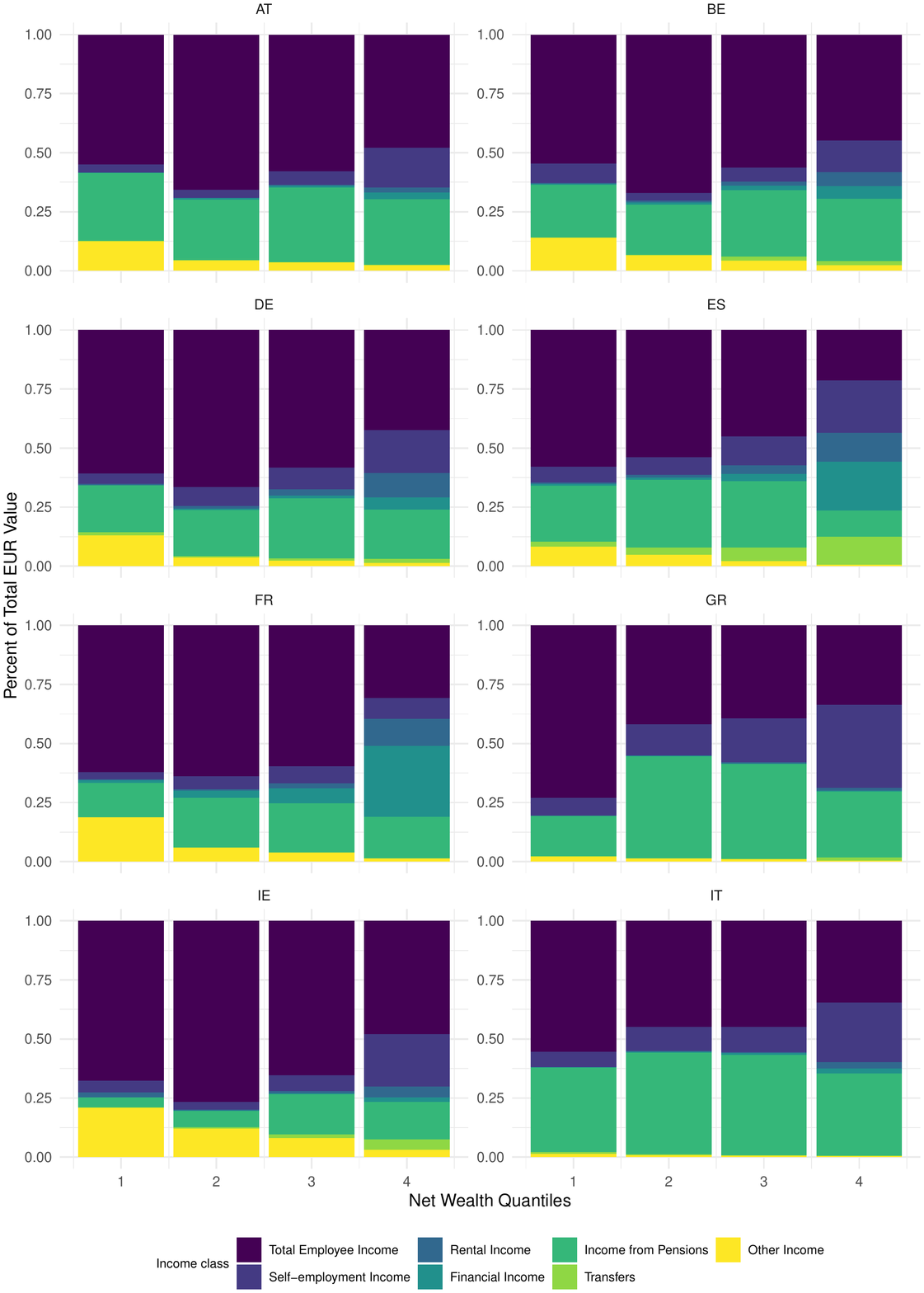}
      \caption{Distribution of types of income across net wealth quantiles.}
      \label{fig:inc_dist}
    \end{minipage}
    \hfill
    \begin{minipage}[t]{.5\textwidth}
      \centering
      \includegraphics[scale=0.5]{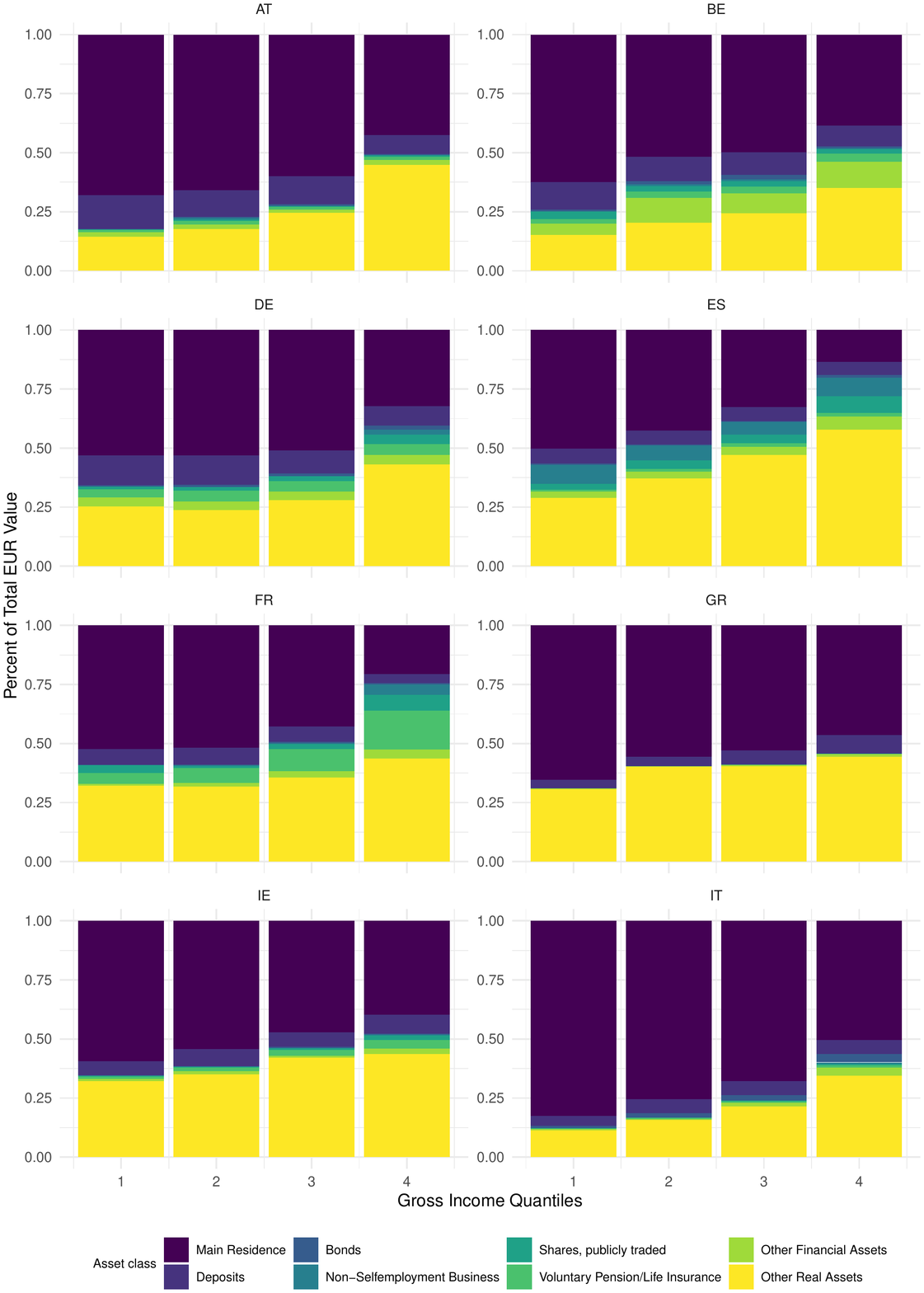}
      \caption{Distribution of net wealth asset classes across income quantiles.}
      \label{fig:nw_dist}
    \end{minipage}
    \hfill
  \end{sidewaysfigure}

Figure \ref{fig:inc_dist} and \ref{fig:nw_dist} give a simple descriptive summary of the distribution of different income components across net wealth quartiles, and the distribution of asset classes across income quartiles, respectively. 
While cross-country differences are clear in these two figures, a few common characteristics stand out.
While the income composition in the fourth quartile is most heterogeneous, by far the most important source of income for most households across all quartiles is employee income.
Income from pensions also plays a significant role in most countries, with Ireland and to some extent France being exceptions here.
As for net wealth, most wealth across all countries and quartiles is made up by real estate, with the household main residence being the most important for lower quartiles and other real estate gaining increasing importance in higher quartiles.
Additionally, Figure \ref{fig:inc_dist} shows kernel density estimates of the joint distribution of income and wealth.
Again, substantial heterogeneity is visible across countries, but all countries indicate a positive association between income and wealth to some degree, as supported by the literature cited in the introduction.\footnote{A straight line with a positive incline would indicate a perfectly positive correlation in these plots.}
It is also noteworthy that clearly the main mass of households in the sample can be found in the lower parts of the distribution, and households with low income and high net wealth (and vice versa) are the exceptions.

\begin{figure}[h]
    \centering
    \includegraphics[width=\textwidth]{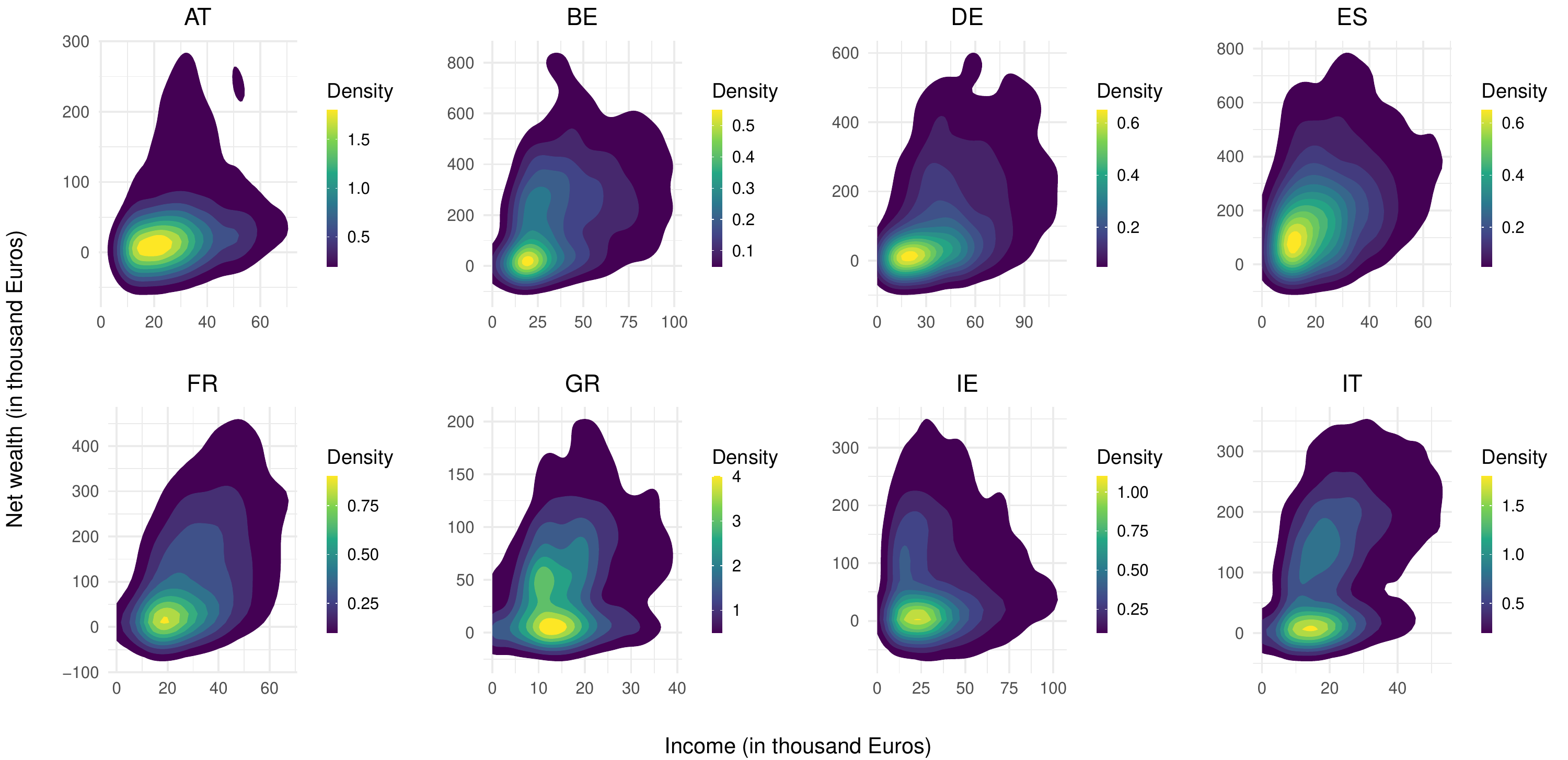}
    \caption{Kernel estimation of the density of the bivariate distribution of income and wealth.}
    \label{fig:inc_wealth}
\end{figure}

\begin{table}[h]
\centering
\caption{Sample values of measures of inequality and correlation and copula-based correlation measures for the initial state and different simulation horizons.}
\label{tab:descriptives}
\resizebox{\columnwidth}{!}{%
\begin{tabular}{l|ccccc|ccc}
  \toprule
  \multicolumn{9}{c}{Sample values}\\
  \midrule
  & \multicolumn{5}{c}{Inequality measures} & \multicolumn{3}{c}{Dependence measures} \\
  \hline
 Country & Gini income & Gini NW$^{\dagger}$ & Gini wealth & Gini debt & biv. Gini & Spearman's $\rho$ & $\lambda_U$ & $\lambda_L$ \\
  \hline
  AT & 0.35 & 0.72 & 0.68 & 0.73 & 0.53 & 0.53 & 0.315 & 0.019 \\ 
  BE & 0.40 & 0.58 & 0.53 & 0.67 & 0.50 & 0.47 & 0.149 & 0.064 \\ 
  DE & 0.46 & 0.68 & 0.64 & 0.68 & 0.57 & 0.57 & 0.303 & 0.000 \\ 
  ES & 0.59 & 0.79 & 0.78 & 0.80 & 0.68 & 0.60 & 0.346 & 0.026 \\ 
  FR & 0.52 & 0.76 & 0.74 & 0.68 & 0.63 & 0.69 & 0.339 & 0.000 \\ 
  GR & 0.37 & 0.58 & 0.53 & 0.64 & 0.49 & 0.37 & 0.111 & 0.093 \\ 
  IE & 0.46 & 0.66 & 0.60 & 0.67 & 0.56 & 0.23 & 0.205 & 0.014 \\ 
  IT & 0.42 & 0.57 & 0.54 & 0.66 & 0.51 & 0.56 & 0.267 & 0.156 \\ 
   \bottomrule
\end{tabular}}
\caption*{\textit{Notes}:  $^{\dagger}$As the Gini coefficient is only defined for positive values, these calculations omit negative values of net wealth. Consequently, values of the Gini coefficient for overall wealth and debt are provided separately as a robustness measure.}
\end{table}

Table \ref{tab:descriptives} summarizes the univariate Gini coefficients for income and net wealth, the bivariate Gini coefficient as well as sample values of the dependence measures, Spearman's $\rho$ and the tail dependence measures $\lambda_U$ and $\lambda_L$.
These simple descriptive calculations give a first impression of the characteristics of the joint distribution of income and wealth.

The Gini index is defined between zero and one, with higher values indicating a more unequal distribution.
As indicated by the Gini index of 0.59, income is most unequally distributed in Spain, followed by France (0.52) and Ireland and Germany (0.46).
Compared to values of the Gini index of income, net wealth is even more unequally distributed. 
Again, Spain and France exhibit the most unequal distributions of net wealth (0.79 and 0.76, respectively), with Austria ranking third with a Gini of 0.72. 

Since net wealth is calculated as the difference between all assets and debts a household has, negative values are possible.
The presence of loans, which are usually available to richer households (due to the fact that they might have better collateral and be more creditworthy), could possibly give a wrong impression of the distribution of assets.
Therefore, Table \ref{tab:descriptives} also includes the Gini coefficients of overall wealth and debt separately.
While Spain, France and Austria still take the top spots in the ranking of the most unequal countries with respect to wealth, the remaining order countries changes slightly. 
Debt is most unequally distributed in Spain (0.80), then Austria (0.73), Germany and France (0.68).

Turning to the bivariate Gini coefficient, measuring inequality in the bivariate distribution of income and net wealth, Spain, not surprisingly, shows most inequality (0.68), followed by France (0.63). 
Germany, with a Gini coefficient of (0.57), takes third place, closely followed by Ireland (0.56).
Note that the bivariate Gini value always lies inbetween the two univariate metrics, with the Gini of income being the lower, and the Gini of net wealth being the upper bound. 

The remaining columns of Table \ref{tab:descriptives} summarize sample values of dependence measures, namely Spearman's $\rho$ and upper and lower dependence measures $\lambda_U$ and $\lambda_L$.
Spearman's $\rho$ is a measure of rank correlation and is defined between $-1$ and $1$.
$\rho=-1$ would imply perfect negative dependence and that the households richest in wealth are poorest in income, $\rho=1$ is perfect positive dependence and implies that households richest in wealth are also the richest in income.
While Spearman's $\rho$ is a useful measure of overall dependence between two variables, dependence in the tails is particularly interesting with respect to income and net wealth.
$\lambda_U$ and $\lambda_L$ as measures of upper and lower tail dependence, respectively, are useful metrics to gain insights on how dependence between income and wealth is structured amongst the richest and poorest households.
Tail dependence measures capture the idea of concordance in the tails of the bivariate distribution.
Income and net wealth can be described to be upper tail dependent if $\lambda_U>0$ and upper tail independent if $\lambda_U=0$, and a similar definition applies for $\lambda_L$.
Overall dependence between income and net wealth is highest in France (0.69), Spain (0.60), Germany (0.57) and Italy (0.56).
Ireland displays the least dependence (0.23), indicating that higher/lower incomes and net wealth, respectively, do not seem to go together as much as in other countries.
While these sample values give a useful first impression about the metrics of interest, inference can give a deeper understanding about how much uncertainty exists around point estimates.

\begin{figure}[h]
\centering
\includegraphics[width=\textwidth]{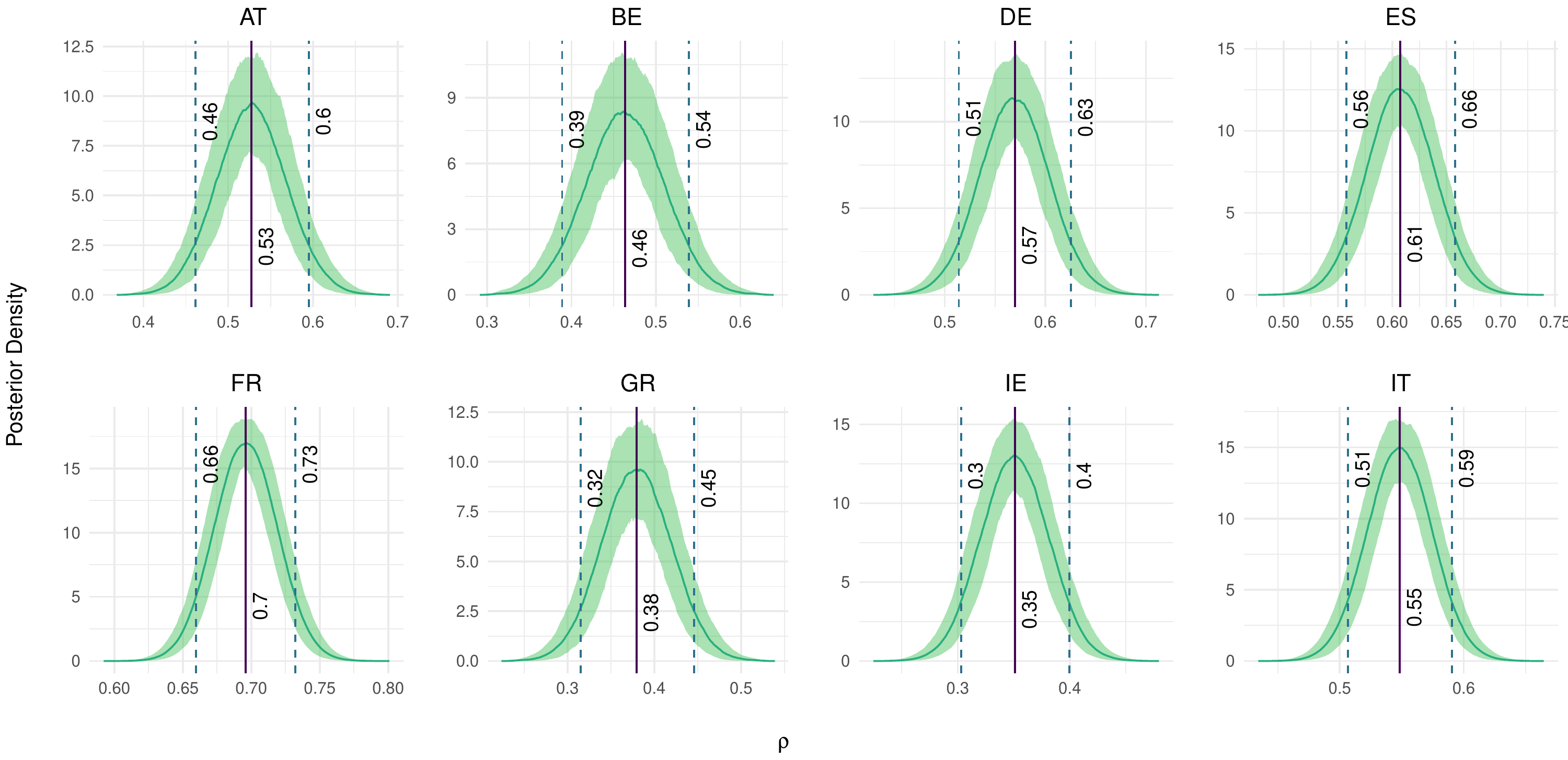}
\caption{Posterior distribution of Spearman's $\rho$ as a overall measure of dependence between income and net wealth.}
\caption*{\footnotesize\textit{Notes}: The vertical solid line indicates the posterior median value over all iterations of the algorithm, the dashed lines mark the 68 percent confidence interval, numbers next to the vertical lines indicate their respective values.}
    \label{fig:rho}
\end{figure}

\begin{figure}[H]
     \centering
     \begin{subfigure}[b]{\textwidth}
        \caption{Upper tail dependence measure $\lambda_U$.}
         \label{fig:lambdaU}
         \includegraphics[width=\textwidth]{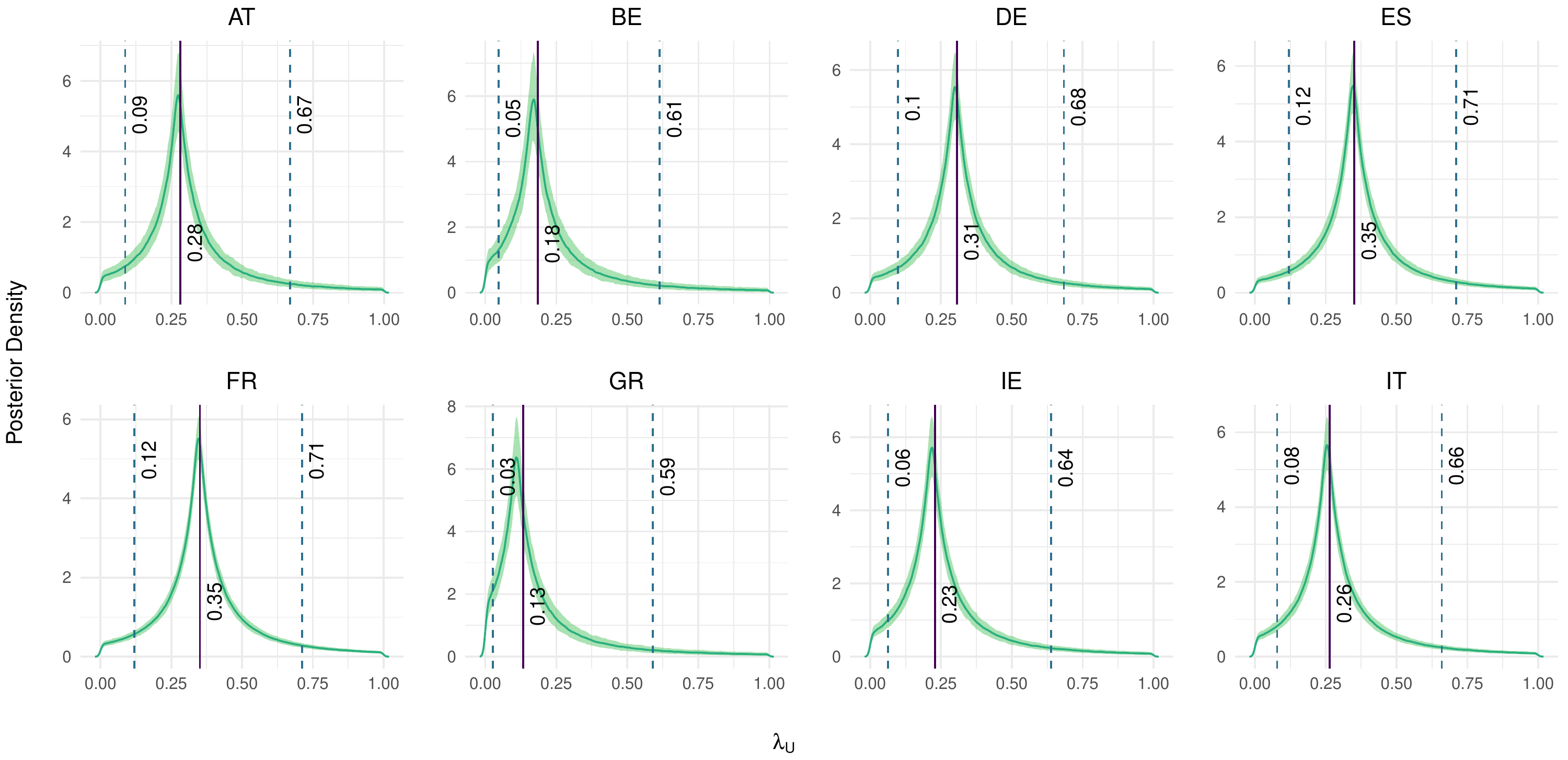}
     \end{subfigure}
     \begin{subfigure}[b]{\textwidth}
        \caption{Lower tail dependence measure $\lambda_L$.}
         \label{fig:lambdaL}
         \includegraphics[width=\textwidth]{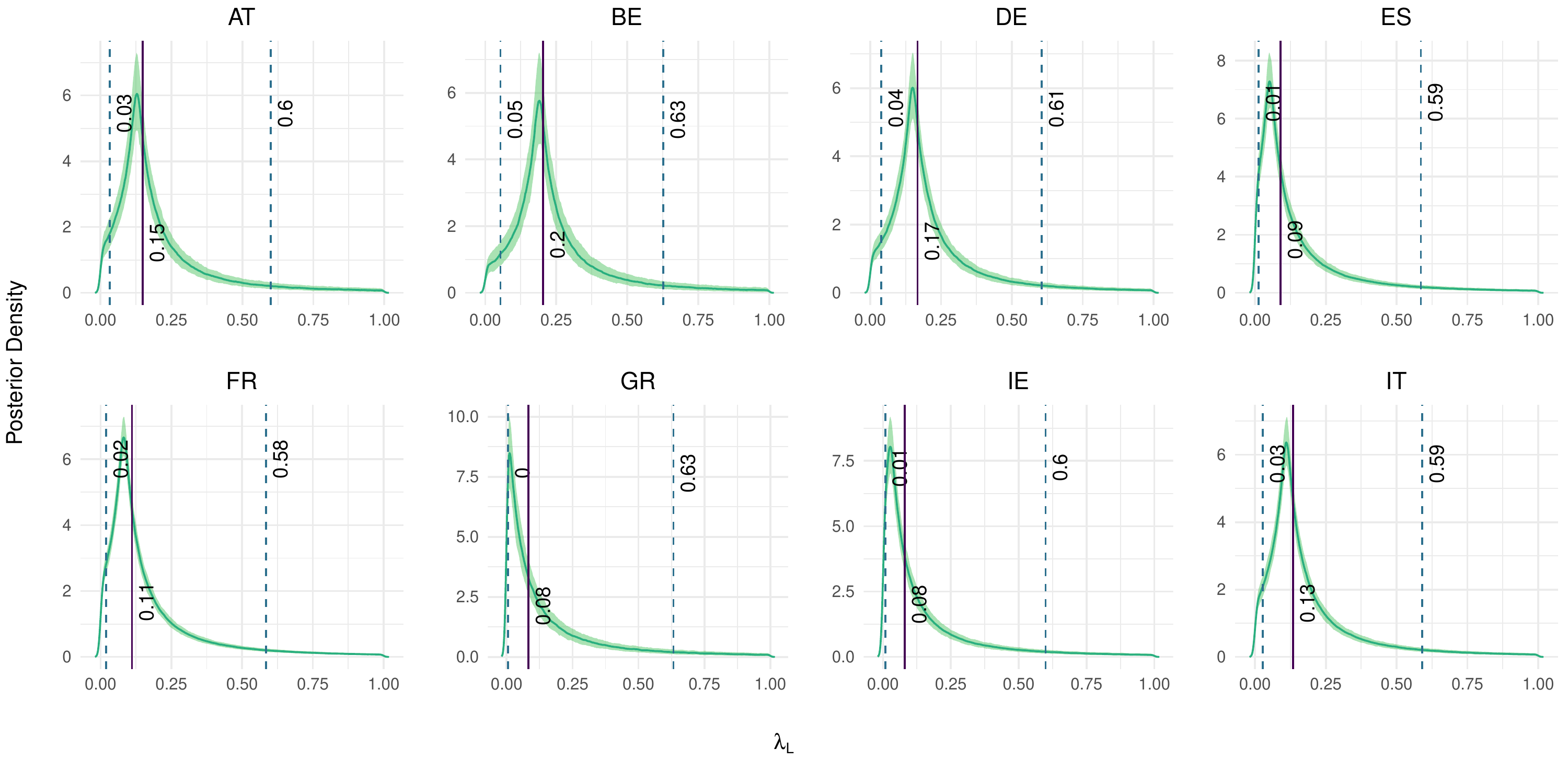}
     \end{subfigure}
        \caption{Posterior distribution of the upper and lower tail dependence measures $\lambda_U$ and $\lambda_L$ respectively, as a measure of tail dependence between income and net wealth.}
        \caption*{\footnotesize\textit{Notes}: The vertical solid line indicates the posterior median value over all iterations of the algorithm, the dashed lines mark the 68 percent confidence interval, numbers next to the vertical lines indicate their respective values.}
        \label{fig:lambdas}
\end{figure}

Figures \ref{fig:rho} and \ref{fig:lambdas} show the copula-based posterior distributions of Spearman's $\rho$ and the tail dependence measures $\lambda_U$ and $\lambda_L$, respectively, for each country.
The shaded area indicates estimation uncertainty stemming from fitting the marginal distributions.
The solid vertical line indicates the median posterior over all copula estimations as well as the marginal parameters.
The dashed lines indicate the 68 percent confidence interval, numbers next to the vertical lines indicate their respective values.

Turning to overall dependence in Figure \ref{fig:rho} first, copula estimates confirm what sample values of Spearman's $\rho$ and the descriptive plots in Figure \ref{fig:inc_wealth} already indicated.
Posterior values of Spearman's $\rho$ are strictly positive in all countries, confirming that income and net wealth do indeed depend positively on each other.
Dependence is weakest in Ireland (0.34) and Greece (0.35) and strongest in France (0.7) and Spain (0.61).
There are several reasons why dependence between income and net wealth can vary so greatly between countries.
First, the lack of public social security nets could explain lower dependence because households have to prepare for possible shocks by having savings on the side.
This would be true for Ireland, which has the lowest expenditure on social protection relative to GDP in this sample of countries, and Greece, which, apart from having high expenditure on pensions, also provides less social protection than other EA countries \citep{eurostatsocialprot}.
Second, if, as \citet{JaenntiSierminskaVanKerm2015} argue, social security payments are conditional on low assets in a means test, households with low income have an incentive to keep their assets low in order to qualify, which would also decrease dependence between income and wealth.

Inspection of Figure \ref{fig:lambdaU} shows that upper tail dependence is present in all countries in the sample, with the 68 percent confidence interval excluding zero and significant posterior mass away from zero.
This a result that was also found by \citet{Chauvel2018IncreasingII} for the USA, who found that there is a trend of convergence of income and wealth at the top over time.
The size of $\lambda_U$ varies between 0.13 in Greece up to 0.35 in Spain and France.
Similarly, posterior estimates of $\lambda_L$ shown in Figure \ref{fig:lambdaL} indicate that income and net wealth are lower tail dependent in all countries except for Greece, where the 68 percent confidence interval includes zero.
The lowest values can be observed in Ireland (0.08) and Spain (0.09), and largest values in Germany (0.17) and Belgium (0.2).
Overall, while there seems to be evidence for some degree of lower tail dependence in most countries, the magnitude of $\lambda_L$ tends to be smaller than $\lambda_U$.
Note, however, that posterior estimates of $\lambda_U$ and $\lambda_L$ are less precise when compared to estimates of Spearman's $\rho$ in Figure \ref{fig:rho}, as the 68 percent confidence interval is much wider.
This is a consequence of the fact that the amount of available data in the tails of the distribution is limited, resulting in larger estimation uncertainty.

Nevertheless, a few conclusions can be drawn from this analysis of dependence between income and net wealth.
First, in line with findings in \citet{JaenttiSierminskaSmeedingOECD2015} and \citet{Chauvel2018IncreasingII}, income and net wealth depend positively on each other in the selected sample of EA countries.
Second, overall dependence varies to quite a large degree between countries, with Greece and Ireland showing the lowest and France and Spain the highest degree of dependence.
Third, evidence for upper tail dependence seems to be stronger than lower tail dependence. 
This indicates that the concordance between households with the most income and net wealth is higher than the concordance in the lower tail of the joint distribution of income and wealth.
One possible explanation for this result is that, as mentioned before, net wealth takes debt into account and households which were able to take out a substantial loan, e.g. a mortgage, need to provide considerable collateral and income in order to ensure repayment.
This would put such a household on the bottom of the net wealth distribution, but place it higher in the distribution of income and, therefore, reduce lower tail dependence.

\subsection{Macroeconomic effects of monetary policy}
\label{sec:macroresults}

\begin{figure}[H]
     \centering
     \begin{subfigure}[b]{\textwidth}
        \caption{Standard monetary policy shock.}
        \label{fig:var_tg}
         \includegraphics[width=0.95\textwidth, trim=10 0 10 10, clip]{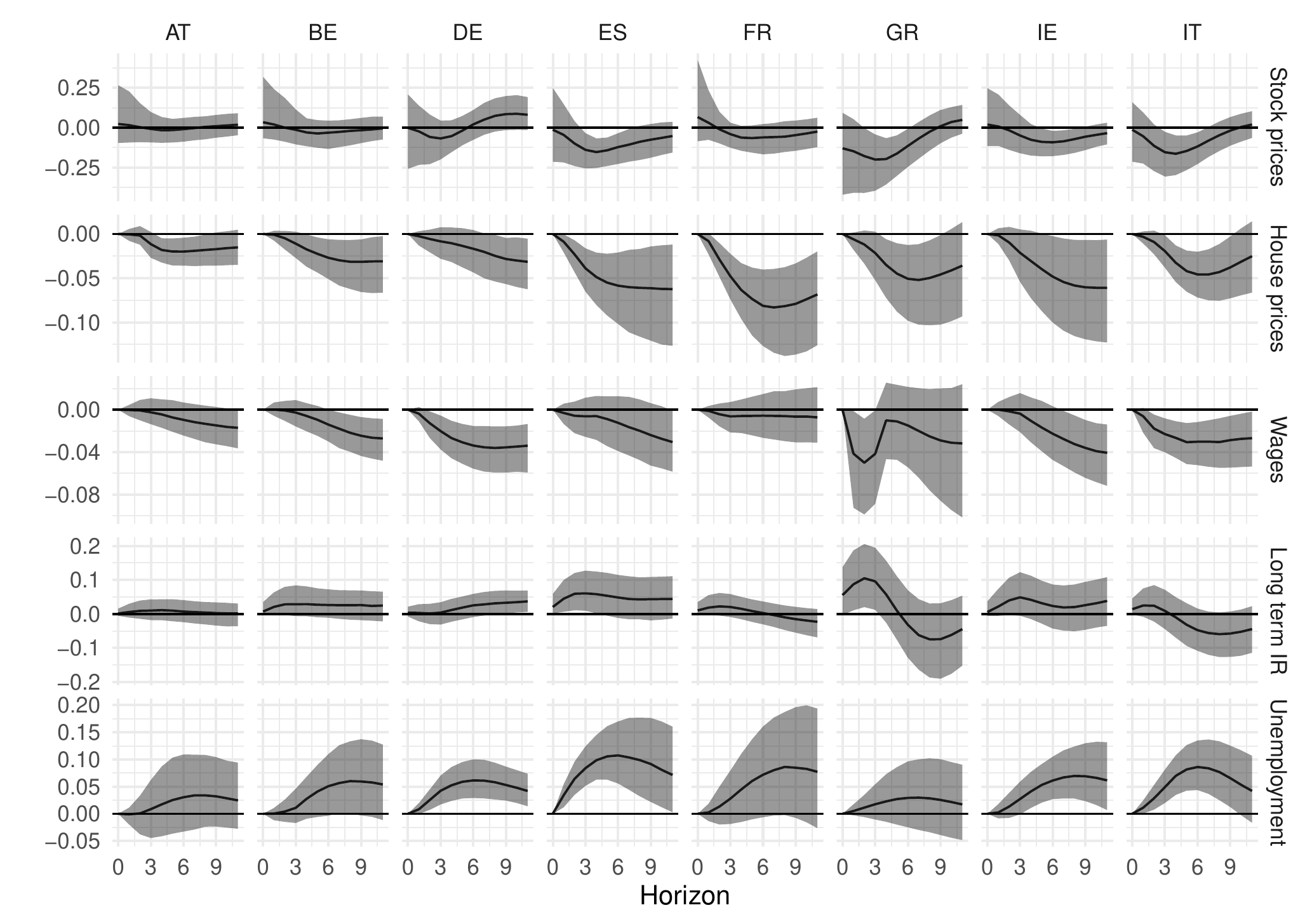}
     \end{subfigure}
     \begin{subfigure}[b]{\textwidth}
        \caption{Quantitative easing shock.}
        \label{fig:var_qe}
         \includegraphics[width=0.95\textwidth, trim=10 0 10 10, clip]{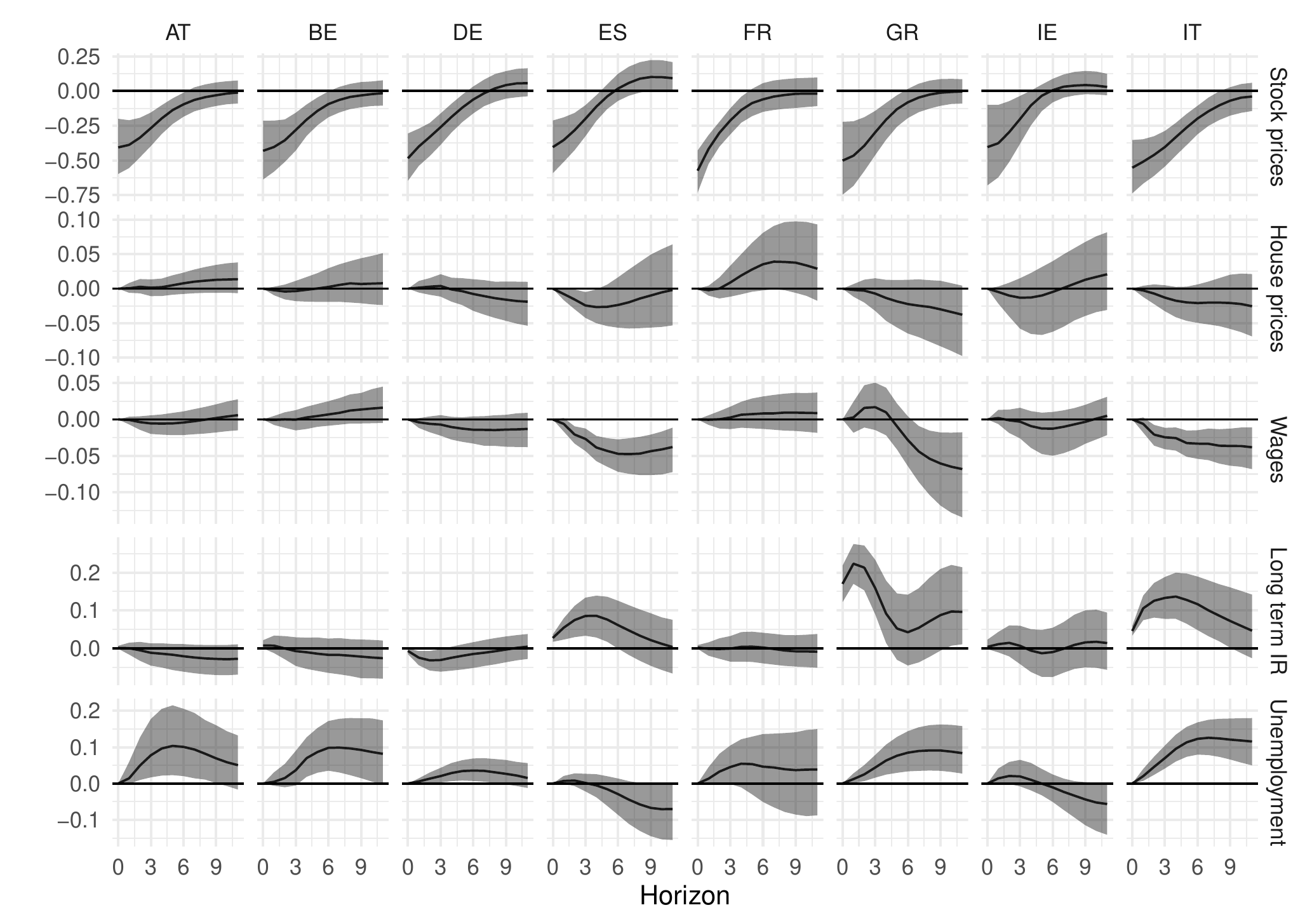}
     \end{subfigure}
        \caption{Impulse response functions from a BVAR to a contractionary one standard deviation standard monetary policy shock and quantitative easing shock.}
        \caption*{\footnotesize\textit{Notes}: Median response alongside the 68 percent posterior credible set. The black horizontal line marks zero.}
        \label{fig:varIRFs}
\end{figure}

The previous section establishes a descriptive overview of the joint distribution of income and net wealth.
Next, I discuss the macroeconomic effects of monetary policy.
Figure \ref{fig:varIRFs} provides IRFs obtained from the BVAR model described in Section \ref{sec:macroMP}.
Figure \ref{fig:var_tg} shows effects of a contractionary one standard deviation shock to the target rate on the variables of interest\footnote{Further results for control variables as well as the variables of interest can be found in the appendix.} and Figure \ref{fig:var_qe} contains effects of a contractionary one standard deviation QE shock.
The columns in the figure collect the different countries, while rows show the responses of the respective variable.\\
While stock prices tend to decrease as a response to a contractionary target rate shock, this effect is only significantly different from zero around the third quarter until the sixth quarter in the southern European countries Spain, Greece and Italy. 
Responses in Germany, France and Ireland show the same dip in stock prices, as well as a slight rebound in the sixth quarter, they are however, not significantly different from zero.
Similarly, responses in Austria and Belgium are virtually flat.
House prices show a much stronger decline in all countries.
The dip in house prices is most pronounced between the third and sixth quarter, with France showing the greatest decrease in house prices, followed by Spain, Ireland and Greece.
Wages show more heterogeneous responses to the target shock.
In Austria, Spain and France, wages do not react in a significant way to the shock.
Wages in Belgium and Ireland drop and are significantly different from zero after the sixth quarter.
Labor compensation in Germany and Italy reacts more quickly and shows significant drops after the first quarter.
Lastly, wages in Greece drop sharply right after the shock but show a strong rebound after four quarters.
Similarly to stock prices, responses of the long term interest rate to a target shock are mostly flat and/or insignificant.
Spain and Greece are the exception with a small significant increase between the second and third quarter.
Lastly, the unemployment rate increases in all countries in response to a contractionary target shock, and while all countries show substantial posterior mass above zero, only Germany, Spain, Ireland and Italy show responses significantly different from zero between the second and tenth quarter.

After a contractionary one standard deviation QE shock, stock prices react much stronger compared to the target shock, with a significant decline until the sixth quarter in all countries.
House prices, on the other hand, react much less and responses show different dynamics over the countries, which, however, stay insignificant over all horizons.
Wages similarly show only dampened responses in most countries.
Exceptions are Spain and Italy, where a drop in wages can be observed on impact and that drop stays significantly different from zero over all horizons, and Greece, where wages first increase insignificantly and then drop after the sixth quarter and this decline is sustained over the remaining horizons.
The long term interest rate hardly reacts to the QE shock in Austria, Belgium, France and Ireland.
In Germany, although very small in size, the long term interest rate decreases significantly between impact and the third quarter.
Spain and Italy show significant decreases, which stay significant a little bit longer in Italy.
In Greece, an even stronger increase in the long term interest rate is visible, which dies down after the sixth quarter.
Unemployment reacts to the contractionary QE shock in accordance with theory in Austria, Belgium, Germany, Greece and Italy, showing significant increases between the first and at least the ninth quarter.
Contrary, no significant effects can be found in Spain, France and Ireland.

\subsection{Effects of monetary policy on bivariate inequality and the distribution of income and net wealth}
\label{sec:descresults_postsim}

\begin{figure}[!h]
     \centering
     \begin{subfigure}[b]{\textwidth}
        \caption{Gini coefficient of income.}
        \label{fig:gini_inc_tg}
         \includegraphics[width=\textwidth]{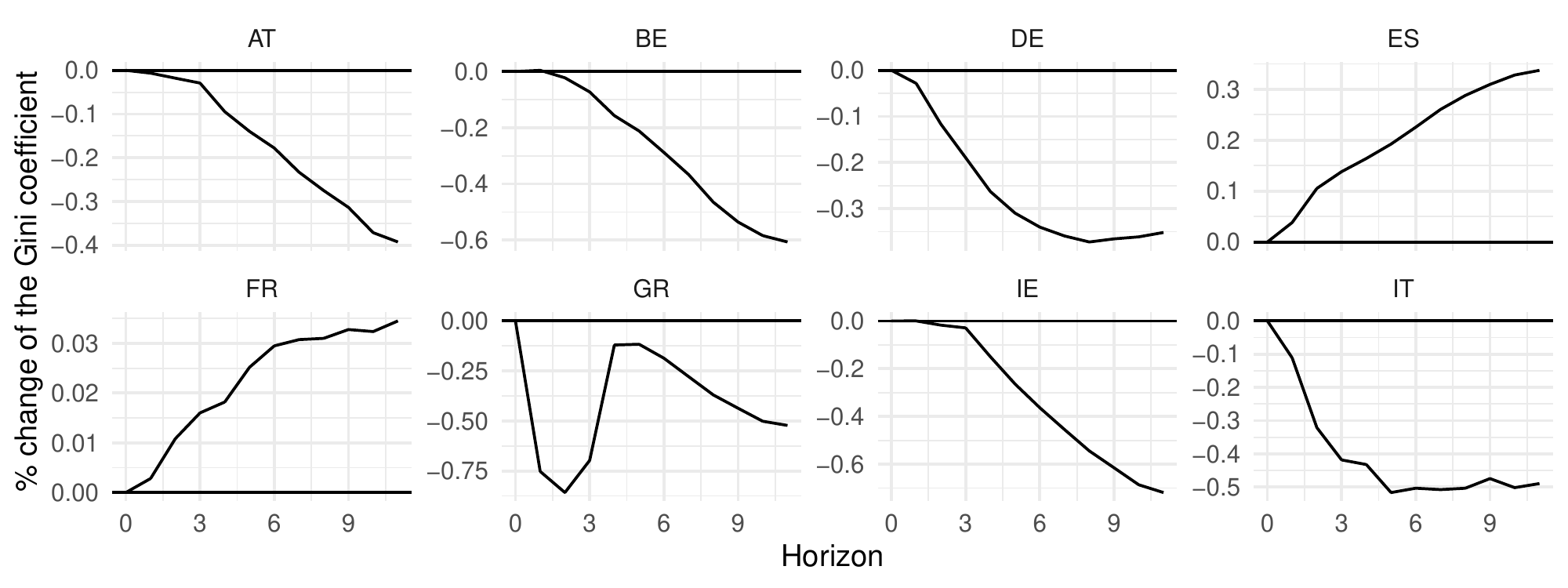}
     \end{subfigure}
     \begin{subfigure}[b]{\textwidth}
          \caption{Gini coefficient of net wealth.}
          \label{fig:gini_nw_tg}
         \includegraphics[width=\textwidth]{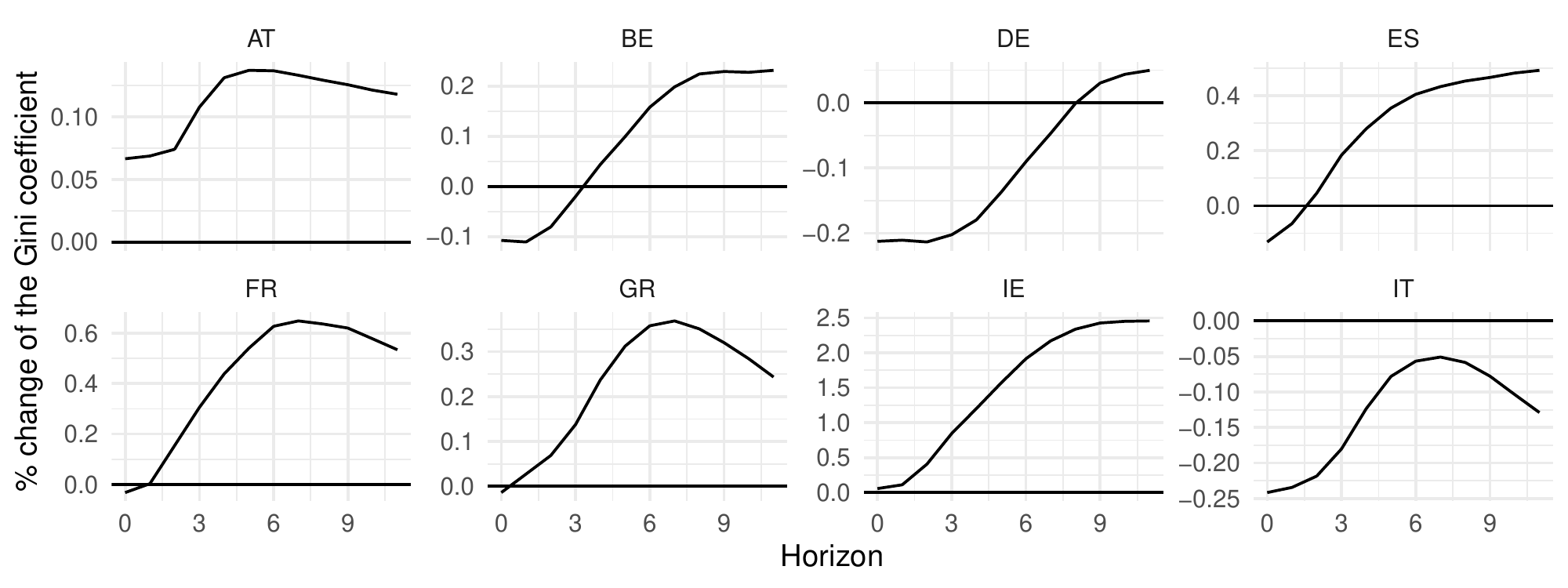}
     \end{subfigure}
          \begin{subfigure}[b]{\textwidth}
          \caption{Bivariate Gini coefficient of income and net wealth.}
          \label{fig:gini_megc_tg}
         \includegraphics[width=\textwidth]{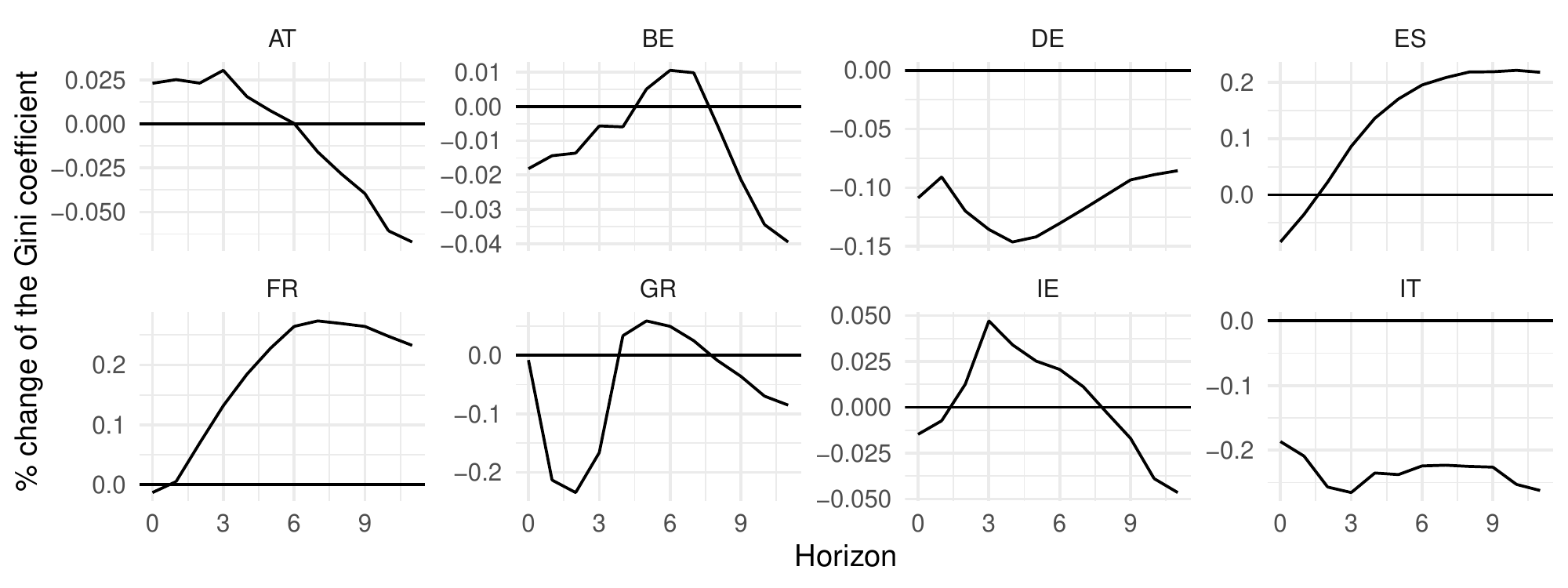}
     \end{subfigure}
      \caption{Simulation results of a contractionary target shock on the Gini coefficient of income, net wealth and the bivariate Gini of both income and net wealth.}
        \caption*{\footnotesize\textit{Notes}: Dynamics are shown in percentage changes relative to the value of the Gini coefficient in the initial state, i.e., the sample value before conducting the simulation. The values shown correspond to the posterior median of VAR responses.}
        \label{fig:gini_tg}
\end{figure}

\begin{figure}[!h]
     \centering
     \begin{subfigure}[b]{\textwidth}
        \caption{Gini coefficient of income.}
        \label{fig:gini_inc_qe}
         \includegraphics[width=\textwidth]{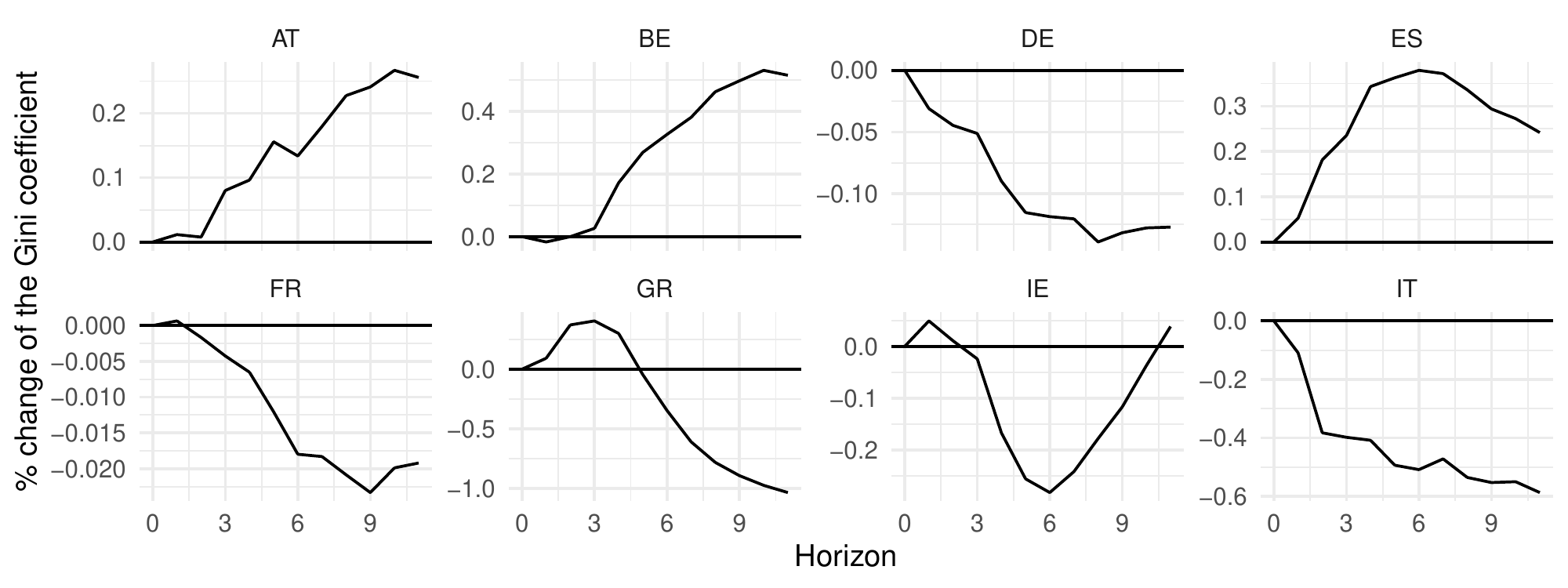}
     \end{subfigure}
     \begin{subfigure}[b]{\textwidth}
          \caption{Gini coefficient of net wealth.}
          \label{fig:gini_nw_qe}
         \includegraphics[width=\textwidth]{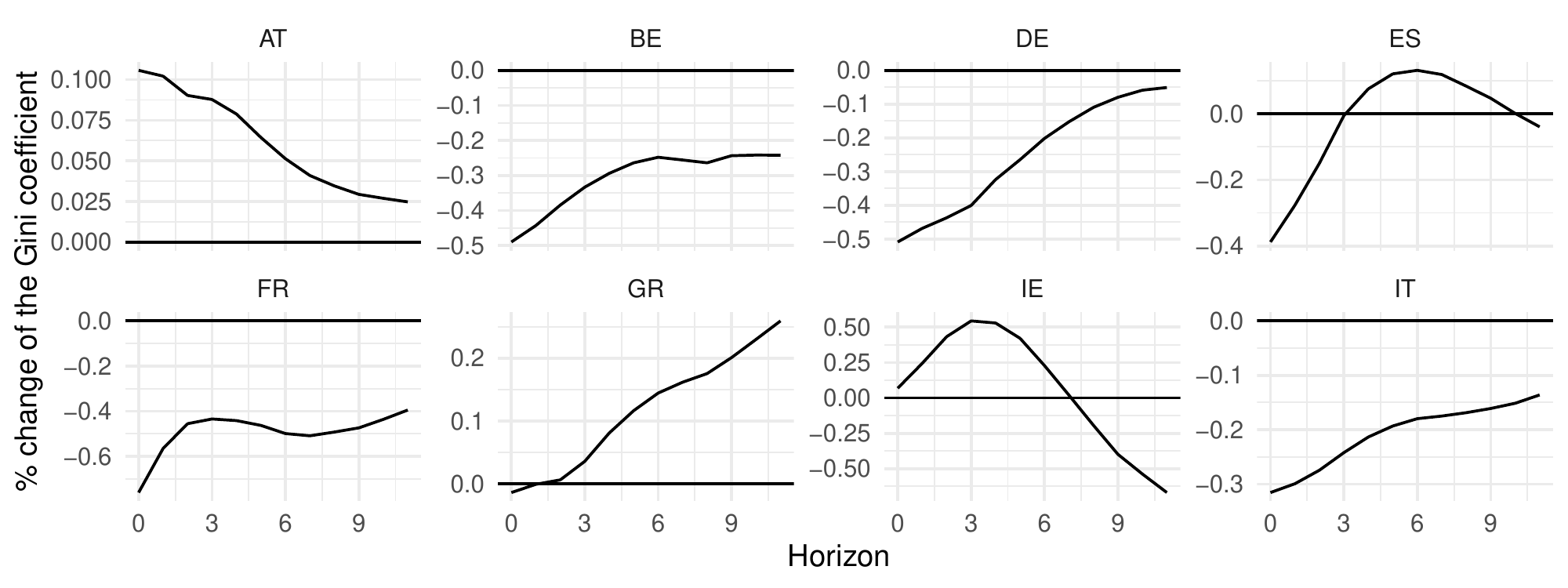}
     \end{subfigure}
          \begin{subfigure}[b]{\textwidth}
          \caption{Bivariate Gini coefficient of income and net wealth.}
          \label{fig:gini_megc_qe}
         \includegraphics[width=\textwidth]{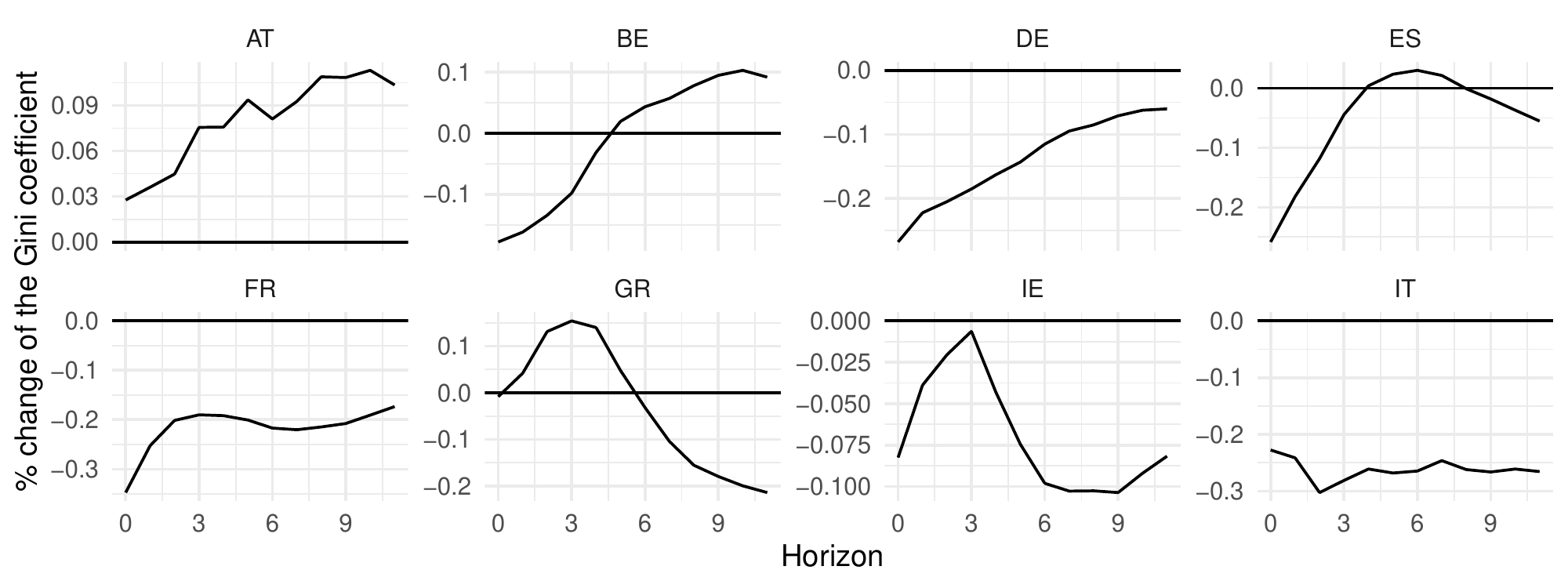}
     \end{subfigure}
        \caption{Simulation results of a contractionary QE shock on the Gini coefficient of income, net wealth and the bivariate Gini of both income and net wealth.}
        \caption*{\footnotesize\textit{Notes}: Dynamics are shown in percentage changes relative to the value of the Gini coefficient in the initial state, i.e., the sample value before conducting the simulation. The values shown correspond to the posterior median of VAR responses.}
        \label{fig:gini_qe}
\end{figure}

This last section of results will provide illustrations of the effects of monetary policy shocks on the distribution of households' income, net wealth and their joint distribution.
Figures \ref{fig:gini_tg} and \ref{fig:gini_qe} provide the dynamic evolution of the Gini coefficients after a contractionary target and QE shock, respectively.
The three panels of the figures show the Gini coefficient of income, net wealth and the bivariate Gini coefficient, which takes both variables into account, and depict the development of the Gini coefficient according to the posterior median of VAR responses shown in Figure \ref{fig:varIRFs}, relative to its initial value.

As shown in Figure \ref{fig:gini_inc_tg}, income inequality as measured by the Gini coefficient reacts in very heterogeneous ways to a contractionary target shock.
This is in line with a previous study by \citet{Guerello2018}, who also found that income inequality reacts very differently in various EA countries. 
In Austria, Belgium, Germany and Ireland, income inequality does not react at all until around the third quarter, after which it starts decreasing.
Income inequality in Italy follows a similar pattern, decreasing immediately after the shock. 
Reductions in the Gini coefficient are overall small and lie between 0.3 percent in Germany and around 0.6 percent in Belgium and Ireland.
Greek income inequality almost perfectly mirrors the wage dynamics in response to the target shock as displayed in Figure \ref{fig:varIRFs}, dropping sharply in the first quarter and rebounding after that, but staying below its initial value.
The Gini coefficient of income in Greece initially drops by 0.75 percent as a response to the contractionary target shock and levels out to a reduction of 0.5 percent after twelve quarters.
Income inequality in France and Spain contrarily rises as a result of a contractionary target shock, albeit to different degrees.
The increase in France is only 0.03 percent, but 0.3 percent in Spain.

Inequality in net wealth behaves in a more homogeneous way, as visible in Figure \ref{fig:gini_nw_tg}.
In Austria and Ireland, the Gini coefficient of net wealth increases immediately after a target shock, while it drops between the first and third quarter in Belgium, Spain, and, to a lesser degree in France and Greece before rising.
In contrast, net wealth inequality in Germany and Italy decreases in response to a target shock, with Germany exhibiting a rebound in the Gini coefficient around the seventh quarter while the Gini of net wealth stays below its initial value for the full horizon.
Another characteristic is a hump shape in the responses of the Gini of net wealth in Austria, France and Greece, with the peak response happening between the fifth and sixth quarter.
In Italy the evolution of the Gini coefficient of net wealth after the contractionary target shock also displays a hump form, but as the Gini here declines, the peak response occurs after the shock.
As with income inequality, changes in net wealth inequality are overall small in size, with the Irish Gini exhibiting by far the greatest changes by 2.5 percent.

Turning to the bivariate Gini of income and net wealth in Figure \ref{fig:gini_megc_tg}, results are again very heterogeneous over the countries in the sample.
In Austria, the Gini coefficient of both income and net wealth first increases slightly until the sixth quarter, after which it decreases. This seems to combine the dynamic of its univariate Gini of income and net wealth, with the latter dominating the first six horizons and the former taking over after that.
Belgium similarly displays a mixed picture, with the bivariate Gini showing and initial decline after the shock, then increasing until its peak response around the sixth quarter and decreasing again after that.
In Ireland the bivariate Gini also increases after a short initial drop and it starts to decrease again after its peak response of 0.05 percent in the third quarter.
Greece shows the same pattern of an initial drop, then increase and decrease, but here it is more pronounced (with a decrease of 0.2 percent) and, as with the income Gini, it is possible to recognize the the dynamic response of wages in the IRF from Figure \ref{fig:varIRFs}.
In Germany the bivariate Gini decreases over the full horizon, with its negative peak response, a decrease of 0.15 percent, around the fourth quarter. 
Italy shows a similar pattern, but with a slightly larger negative peak response of 0.27 percent around the third quarter.
These responses are not surprising, given that both the Gini of income and net wealth decreased in these two countries as a response to the contractionary target shock.
Similarly, Spain and France, where the Gini coefficient of income and net wealth both increased after the target shock, show an increase in the bivariate Gini coefficient after a short initial decrease. 
In Spain, the response peaks at about 0.22 percent around the seventh quarter, and at about 0.27 percent at the same time horizon.

The effects of a contractionary QE shock on income inequality are displayed in Figure \ref{fig:gini_inc_qe}.
The Gini coefficient of income increases in Austria, Belgium and Spain. 
Increases peak at around 0.27 percent in Austria, 0.47 percent in Belgium and 0.37 in Spain.
Income inequality in Ireland experiences a short increase (until the second quarter) and Greece (until the fifth quarter) before decreasing.
In Germany, France and Italy, income inequality decreases over the horizon, with a negative peak response of 0.18 percent in Germany at the eighth quarter, 0.02 percent in France at the ninth quarter, and the largest negative peak response of almost 0.6 percent in Italy around the eleventh quarter.

Inequality in net wealth, depicted in Figure \ref{fig:gini_nw_qe}, decreases in Belgium, Germany, France and Italy over the whole horizon.
The decreases range from 0.3 in Italy to about 0.65 percent in France.
In Spain, it decreases immediately after the shock and then returns to its initial value around the third quarter, after which it experiences a little rebound between the third and ninth quarter.
Similarly, the Gini coefficient of net wealth increases comparatively strongly after the shock in Ireland (more than 0.5 percent), returns to its initial value around the seventh quarter and falls after that.
In Austria and Greece, net wealth inequality increases after the QE shock, although the dynamic is different. 
In Austria, the peak response of 0.1 percent occurs right after the shock after which the Gini slowly moves back to its initial value. 
In Greece however, net wealth inequality continually rises after the shock, up to a 0.25 percent increase at the eleventh quarter.

Similarly to the target shock, the joint measure of the bivariate Gini in Figure \ref{fig:gini_megc_qe} shows very mixed results.
In accordance to an increased Gini of income and net wealth, the bivariate Gini increases in Austria, although the upward trending dynamic from the income Gini seems to dominate.
Similarly, but with reversed sign, overall inequality decreases in Germany, France and Italy over the whole horizon.
In Belgium and Spain, the differing dynamics in income and net wealth inequality result in an initial decrease in the bivariate Gini, which returns to its initial value around the fifth quarter and increases further after that.
While the bivariate Gini keeps increasing after that in Belgium, it falls again in Spain.
Conversely, the contractionary QE shock initially increases the bivariate Gini in Greece until its peak response of 0.15 percent a the third quarter, after which it starts to fall.
Ireland displays a similar dynamic, yet here the bivariate Gini decreases over the whole horizon.

In conclusion, a one standard deviation contractionary target shock tends to decrease inequality in income and increase inequality in net wealth in most countries.
The exceptions are Spain and France, which display increases in income inequality, and Germany and Italy, which experience less inequality in net wealth following the shock. 
These contrary effects result in mixed results for bivariate inequality in most countries, but a clear increase in Spain and France, and a distinct decrease in Germany and Italy.
A one standard deviation contractionary QE shock tends to decrease income inequality in Germany, France, Italy and, over most of the horizon, Ireland, but increases income inequality in Austria, Belgium and Spain. 
Net wealth inequality tends to decrease after the QE shock in Belgium, Germany, France and Italy, but increases in Austria and Greece, while Spain and Ireland show mixed results depending on the horizon.
Overall, bivariate inequality is increased by the QE shock in Austria, but decreased in Germany, France, Ireland and Italy.
The response of bivariate inequality in Belgium, Spain and Greece again depends on the horizon.
This mixed results are in line with findings by \citet{lenza2021} and \citet{Guerello2018}.
As discussed in the introduction of this paper, the existing literature does not agree on the final effect monetary policy has on income and wealth inequality.
Differences can stem from diverse portfolio allocations or households' different exposure to the financial market \citep{Guerello2018} or, more generally, to various transmission channels of monetary policy \citep{andersen2022monetary}.
The state of the business cycle may also impact the effects of monetary policy on inequality, as suggested by results in \citet{ofarrell2016} and \citet{furceri2018}.

\begin{figure}[!h]
     \centering
         \includegraphics[width=\textwidth]{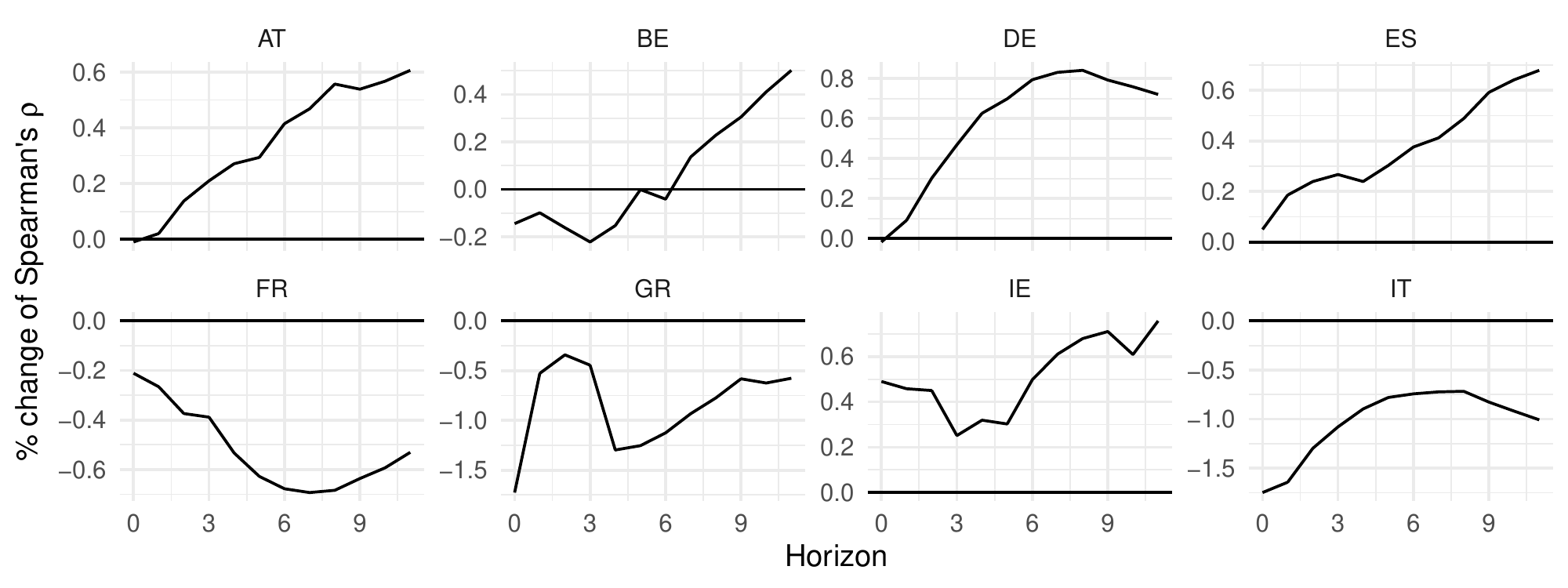}
        \caption{Simulation results of a one standard deviation contractionary target shock on dependence measure Spearman's $\rho$ of income and net wealth.}
        \caption*{\footnotesize\textit{Notes}: Dynamics are shown in percentage changes relative to values in the initial state, i.e., the copula-based values before conducting the simulation. The values shown correspond to the posterior median of VAR responses.}
        \label{fig:dependence_postsim_tg}
\end{figure}

\begin{figure}[!h]
     \centering
         \includegraphics[width=\textwidth]{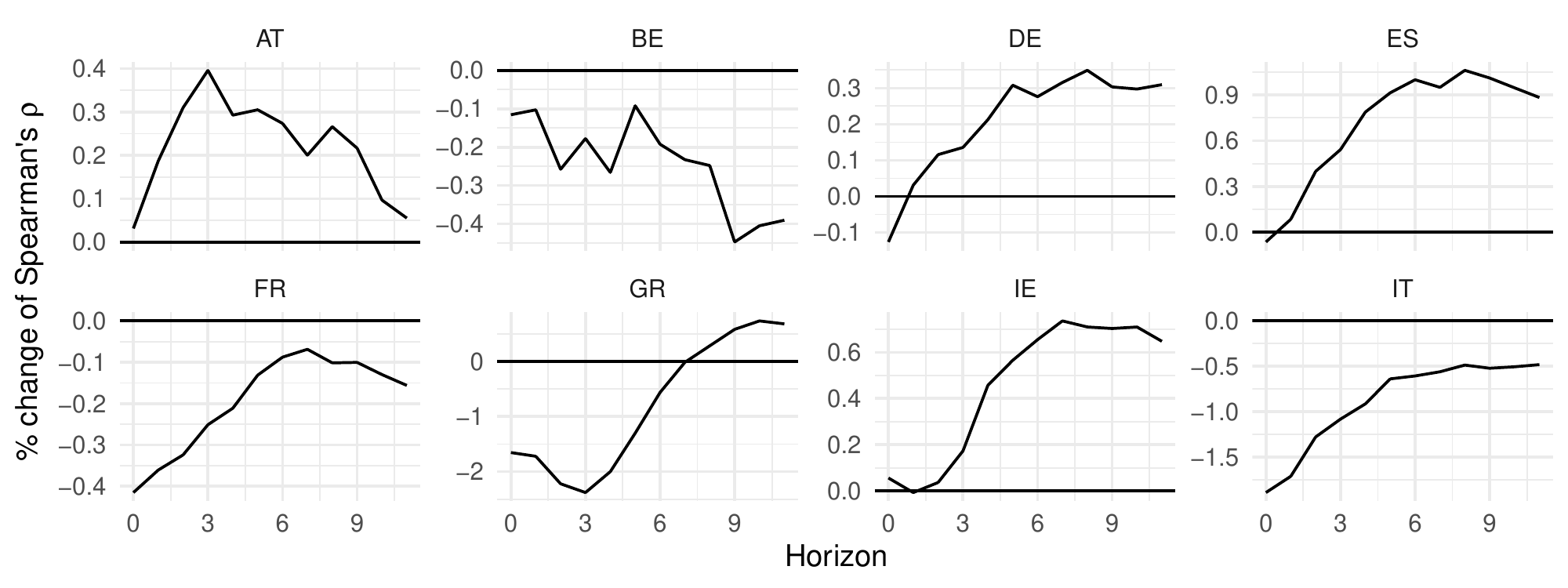}
        \caption{Simulation results of a one standard deviation contractionary QE shock on dependence measure Spearman's $\rho$ of income and net wealth.}
        \caption*{\footnotesize\textit{Notes}: Dynamics are shown in percentage changes relative to values in the initial state, i.e., the copula-based values before conducting the simulation. The values shown correspond to the posterior median of VAR responses.}
        \label{fig:dependence_postsim_qe}
\end{figure}

The evolution of dependence between income and net wealth as illustrated in Figures \ref{fig:dependence_postsim_tg} and \ref{fig:dependence_postsim_qe} conclude the empirical findings of this paper.
Changes in the dependence measure are again presented in percentage changes relative to their initial value and represent the posterior median of Spearman's $\rho$.
Posterior values of Spearman's $\rho$, depicted in Figure \ref{fig:dependence_postsim_tg}, indicate that overall dependence between income and net wealth tends to increase up as a response to a contractionary target shock in Austria, Germany, Spain and Ireland.
In contrast, dependence between income and net wealth decreases in France, Greece and Italy.
Belgium provides a mixed picture, with an initial decrease in Spearman's $\rho$ and increases after the sixth quarter.

Interestingly, effects of a one standard deviation contractionary QE shock seem to translate into a very similar dynamics of overall dependence compared to the target shock, as visible in Figure \ref{fig:dependence_postsim_qe}.
Again, Spearman's $\rho$ increases in Austria, Germany, Spain and Ireland.
Contrarily, Belgium, France and Italy exhibit a drop in overall dependence between income and net wealth, while in Greece, overall dependence first drops and then rebounds after the third quarter.

In conclusion, a contractionary target shock will increase overall dependence between income and net wealth in most countries, with France, Greece and Italy being the exception.
This is interesting from a policy perspective as households in countries which experience a rise in dependence between income and net wealth might be less able to smooth consumption with savings in the face of decreasing income, especially in those countries, which had a higher association between income and net wealth to start with, like Germany, Spain or Austria.
The responses of dependence between income and net wealth to a contractionary QE shock are more mixed, with rising values of Spearman's $\rho$ in Austria, Germany, Spain and Ireland, and decreasing numbers in Belgium, France, Italy, and, over most of the horizon, Greece.
Again, increases in association between income and net wealth affect some of the countries which started out with higher overall dependence, such as Austria, Germany and Spain, which could affect households' capability to smooth consumption.

\subsection{Determinants of inequality and the dependence between income and net wealth}
\label{sec:resultscountrychar} 

 This last set of results tries to shed some light on the differences observed in the responses described in section \ref{sec:macroresults}.
 To this end, I conduct a very simple exploratory pairwise regression between peak responses of the Gini of income, net wealth, the bivariate Gini and the dependence between the two variables and a set of possible explanatory variables.
 Similar to the results in section \ref{sec:descresults_postsim}, peak responses are calculated relative to their initial values and thus represent the peak change of the respective metric after a shock.

 The full set of countries is considered as a cross-section for this analysis. 
 All possible explanatory variables are aggregated from HFCS data in order to keep data sources as consistent as possible, and include both demographic and economic information about each country.\footnote{The possible characteristics include the unemployment rate, average income from pensions (Avg. pension income),
 average unemployment benefits (Avg. UB) and average social transfers (Avg. ST, as proxies for the strength of the welfare state), average household income (Avg. income), average household net wealth (Avg. NW), average household debt (Avg. debt), share of people with occupational pension plan (\% occupational pension), share of people with voluntary pension schemes (\% voluntary pension), share of people with public pension plans (\% public pension), share of people who are self-employed (\% self-employed), average financial assets (Avg. financial wealth), share of households that hold investment in some business (\% business investment),
 share of households that own any financial wealth (\% financial wealth), share of households which own their main residence (\% home owners), share of people who finished tertiary education (\% tertiary education), average age (Avg. age), share of retired people (\% retired), average number of dependent children per household (Number of children), share of single persons in the economy (\% single), average household size (Avg. HH size) and the initial values of the dependence measure Spearman's $\rho$ (Dependence $t=0$), the bivariate Gini of income and net wealth (Bivariate Gini $t=0$) and the Gini of net wealth (NW Gini $t=0$) and income (Income Gini $t=0$), respectively.}

\begin{figure}[H]
     \centering
        \includegraphics[width=\textwidth]{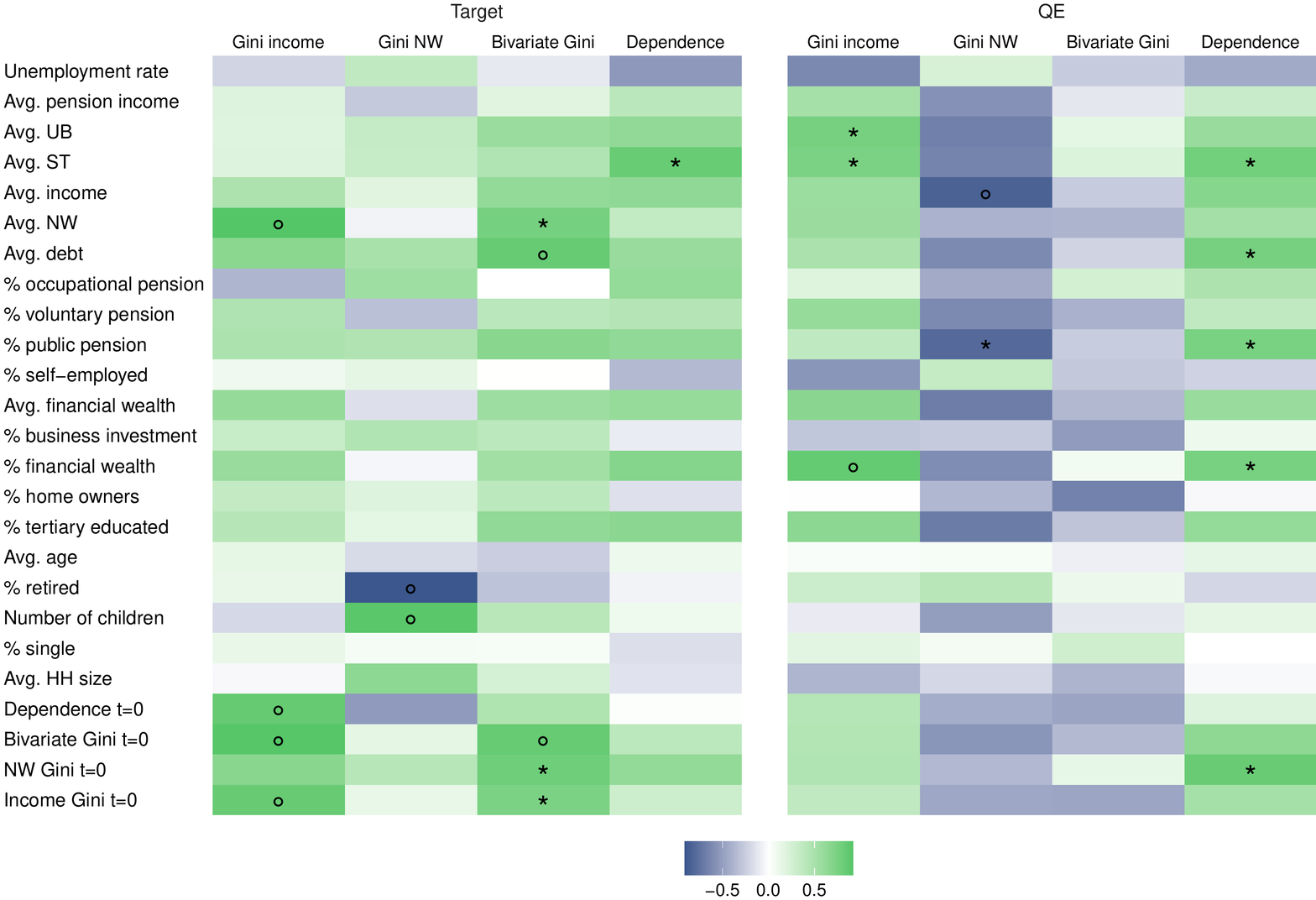}
        \caption{Pairwise exploratory regression between metrics of the distribution of income and net wealth and possible explanatory variables.}
        \caption*{\footnotesize\textit{Notes}: The dependent variables -- Gini of income (Gini income), of net wealth (Gini NW), of both income and net wealth (Bivariate Gini), and Spearman's $\rho$ of income and net wealth (Dependence) --  are depicted on the horizontal axis for both a target and an quantitative easing (QE) shock. Explanatory variables are listed on the vertical axis: unemployment rate, average pension income (Avg. pension income), average level of unemployment benefits (Avg. UB), the average level of social transfers (Avg. ST), the average level of income (Avg. income), the average level of net wealth (Avg. NW), the average level of debt (Avg. debt), the share of people with an occupational pension plan (\% occupational pension), the share of people with a voluntary pension scheme (\% voluntary pension), the share of people who have a public pension plan (\% public pension), the share of people in the population who are self-employed (\% self-employed), average financial wealth (Avg. financial wealth), the share of households that hold some business investment (\% business investment), the share of households with financial wealth (\% financial wealth), the share of households that own their main residence (\% home owners), share of people with a tertiary education (\% tertiary educated), average age (Avg. age), the share of people in the economy who are already retired (\% retired), the average number of children in a household (Number of children), the share of single people (\% single), average household size (Avg. HH size), the initial level of Spearman's $\rho$ (Dependence $t=0$), the initial level of inequality in terms of both income and net wealth (Bivariate Gini $t=0$), the initial level of net wealth inequality (NW Gini $t=0$) and the initial level of income inequality (Income Gini $t=0$). * indicates the $\alpha=0.05$ significance level, ° indicates significance at the level of $\alpha=0.01$.}
        \label{fig:peakresponse_exp}
\end{figure}

 Figure \ref{fig:peakresponse_exp} contains the regression results. Color intensity shows a stronger association; positive coefficients are displayed in shades of green, negative coefficients in shades of blue.
 Statistically significant associations are marked with ° or *, for the $\alpha=0.01$ and $\alpha=0.05$ significance level, respectively.
 All variables in the regressions are standardized to have zero mean and a standard deviation of one, therefore, coefficients refer to how many standard deviations the dependent variables changes after a one standard deviation increase in the explanatory variable.
 Note that this simple regression can merely show an association between two variables and must not be interpreted causally.

A first glance at figure \ref{fig:peakresponse_exp}, which contains responses after a target shock on the left and responses to a QE shock on the right, shows the set of variables with some kind of explanatory power depends on the metric and type of shock in question.
For the Gini coefficient of income, the regression indicates that average net wealth of households and the initial levels of dependence, the bivariate Gini and its own initial value play seem to play a role in the transmission of a target shock.
For all these explanatory variables, the higher the increase in the independent variable is, the stronger the Gini coefficient of income reacts to a change in the target interest rate.
One can conjecture that some threshold effects are at play here, where after a certain level of inequality is reached, dynamics further accelerate income inequality.
Moreover, as both average net wealth, the dependence between income and net wealth and the bivariate Gini show a positive connection with changes in the income Gini, it is possible that higher levels of net wealth and bivariate inequality influence the reaction of income inequality via the dependence between two different household resources.
As for inequality in net wealth, after a target shock, two demographic characteristics seem to influence the change in the Gini coefficient of net wealth.
The higher the share of retired people in a country, the weaker the response in a country's Gini of net wealth tends to be. 
The opposite is true for the average number of children in a household; the more dependent children there are in a household, the stronger the change in the Gini of net wealth after a target shock.
These dynamics might reflect the fact that households' portfolio composition changes over the course of a lifetime.
Figure \ref{fig:agegroups} in the appendix shows the distribution of different income components and net wealth assets over different age groups.
While the household main residence as a wealth asset maintains it relative importance once a household acquires it between the age of 35 and 50, other real assets tend to decrease in both relative and absolute terms after the age of 50 and other assets, such as voluntary pensions or life insurance, shares or deposits increase. This is true for most, but not all countries.
The reaction of the bivariate Gini seems to be solely determined by distributional characteristics.
The higher the initial level of income and net wealth inequality, both individually as well as jointly, the stronger the reaction of the bivariate Gini after a target shock.
Furthermore, the average level of net wealth and debt in a country also show a positive association with the change in the bivariate Gini.
As for the dependence between income and net wealth, only the average amount of social transfers seems to influence the reaction of Spearman's $\rho$ after a target shock, the more generous social transfers are, the stronger the change in dependence between income and net wealth after a target shock.

When a QE shock occurs, other variables show an association with the reaction of the distribution of income and net wealth.
The average level of unemployment benefits and social transfers, both variables which are included in a household's income (if eligible), show a positive association with the change in income inequality.
This is a puzzling result, as one could argue that higher levels so social transfers and unemployment benefits would dampen any effects QE could have on income inequality.
Furthermore, the higher the share of households which hold some kind of financial wealth, the stronger the income Gini reacts to a QE shock.
As more households are exposed to the financial markets, the distribution of their income can also be affected by asset purchasing programs.
As for inequality in net wealth, the average level of income and the share of public pension coverage in a country have a negative association with the change in the Gini of net wealth after a QE shock.
One possible explanation for this negative association between public pension coverage and net wealth inequality is, that as public pension coverage increases, households have to rely less on other forms of investment as savings for old age. 
These alternative forms of savings might be mores susceptible to changes in the ECB's QE policy.
While there seems to be no association between the explanatory variables and the bivariate Gini in the case of a QE transmission, several variables show an association with the reaction of Spearman's $\rho$.
Increases in the average amount social transfers and debt, the share of people in a country, which have a public pension plan or financial wealth and the initial level of net wealth inequality all exhibit, are indicated to lead to stronger changes in the dependence between income and net wealth.

A full causal examination of determinants of an economy's aggregate responses to a monetary policy shock is beyond the scope of this paper and will be left to future research.
There is, however, some existing literature, which might further bring insights to this question.
\citet{ampudia2018}, for example, show that the aggregate demand response to a monetary policy depends on the fraction of ''hand-to-mouth'' households in an economy, which are households with large spending commitments relative to their income or liquid assets, as they account for the biggest share of consumption response to monetary policy shocks.
These households tend to be very sensitive in their discretionary spending to small and temporary changes in income.
Since these households have a larger marginal propensity to consume than others, the effect of monetary policy on their incomes is especially important for the transmission of monetary policy to the real economy.
Therefore, policy that mostly affects these hand-to-mouth households will have different aggregate effects than policy that affects other parts of the joint distribution of income and wealth.
Similarly, \citet{amir2021does} find that the sensitivity of households' employment status to changes in monetary policy depend demographic characteristics like education or family status, i.e., less educated individuals or single males seem to be much more affected by monetary policy shocks than other demographic groups.
Furthermore, \citet{voinea2018impact} find for Romania that the household debt and income seems to play a role in the effective transmission of monetary policy, which seems to corroborate the findings above.

\section{Closing remarks}
\label{sec:conclusion}
This paper measures the effects of a contractionary monetary policy shock, both conventional and unconventional, on the joint distribution of income and net wealth in terms of their dependence on each other in a selected sample EA countries.
To this end, I estimate a semiparametric copula model in order to establish measures of overall and tail dependence and calculate a bivariate Gini coefficient using HFCS household data.
Effects of monetary policy are evaluated on an aggregate using a BVAR, and, in a second step, mapped into detailed data on income and net wealth.
By re-calculating dependence and inequality measures with the data obtained from this microsimulation, I am able to discuss how monetary policy affects those metrics.

The results of this paper suggest that contractionary conventional monetary policy tends to increase inequality in terms of income and net wealth in most EA countries, which is mostly driven by dynamics in wealth inequality.
Furthermore, a contractionary target shock reinforces the positive association between income and net wealth, which can be found in all countries in the sample.
Effects of conventional monetary policy on tail dependence is mixed, and while income and net wealth are lower tail dependent in all EA countries, no common dynamics emerge as a response to a contractionary shock.
Besides these common trends, substantial heterogeneity in the effects of monetary policy can be found, and it is worth mentioning that results across individual countries might differ from EA aggregates.

Responses in inequality to a contractionary QE shock are similarly mixed, decreasing overall inequality in some countries, while increasing it in others.
In contrast, it seems clear that a contractionary QE shock increases dependence between income and net wealth, similarly to a target shock.
Additionally, effects of conventional and unconventional monetary policy both deliver mixed results when it comes to their effect on tail dependence.

In summary, while further research is necessary to identify the reasons for such heterogenous effects of monetary policy, this papers succeeds in establishing a wide variety of different effects of monetary policy on the distribution of income and net wealth.
From a policy perspective, it is highly relevant to be aware of the various ways that monetary policy may affect inequality, both uni- and bivariate, and the dependence of income and net wealth.
This is true both in terms of a time and a geographical dimension.

\small\setstretch{0.8}\addcontentsline{toc}{section}{References}
\bibliographystyle{custom.bst}
\bibliography{lit}

\clearpage
\begin{appendices}\crefalias{section}{appsec}
\begin{center}
{\LARGE\textbf{Appendix}}
\end{center}


\setcounter{equation}{0}
\renewcommand\theequation{A.\arabic{equation}}
\section{Macroeconomic data}
\begin{table*}[h]
\begin{center}
  \caption{Data sources and transformations.}
  \label{tab:dataVAR}
\begin{small}
\begin{threeparttable}
\begin{tabular*}{\textwidth}{@{\extracolsep{\fill}} llll}
  \toprule
  \textbf{Variable} & \textbf{Description} & \textbf{Source} & \textbf{Trans.}\\ 
  \midrule
  & \multicolumn{3}{l}{\textit{Variables for microsimulation}} \\
  \cmidrule{2-4}
  \texttt{DJ50} & Euro Stoxx 50, price index, quarterly mean & SDW & $100\cdot$log$(x)$\\
  \texttt{HP}  & Analytical house prices indicators & OECD & $100\cdot$log$(x)$ \\
  \texttt{LCOMP} &  Compensation of employees (current prices) divided  & OECD & $100\cdot$log$(x)$\\
  & by employment (in persons) & & \\ 
  \texttt{LT-IR} & Euro area 10-year Government Benchmark bond yield, & SDW & \\ 
  & quarterly mean & & \\ 
  \texttt{UNEMP} & Unemployment rate, s.a. & OECD & $100\cdot$log$(x)$\\
  \midrule
  & \multicolumn{3}{l}{\textit{Additional control variables}} \\
  \cmidrule{2-4}
  \texttt{EA-spread} & Euro area government bond spread, own calculations & SDW &\\ 
  & based on \texttt{LT-IR}& & \\ 
  \texttt{GDP} & Gross domestic product & Eurostat & $100\cdot$log$(x)$\\
  \texttt{HICP} & Harmonized consumer price index, mquarterly mean, s.a. & Eurostat & $100\cdot$log$(x)$\\
  \texttt{ST-IR} & Eonia 3-month rate, quarterly mean & SDW & \\ 
\bottomrule
\end{tabular*}
\begin{tablenotes}[para,flushleft]
\scriptsize{\textit{Notes}: Column \textit{Trans.} indicates the transformation applied to the respective series $x$. OECD indicates the database maintained by the Organisation for Economic Co-operation and Development (OECD, \href{https://data.oecd.org}{data.oecd.org}), Eurostat is the database provided by the statistical office of the European Union (\href{https://ec.europa.eu/eurostat/data/database}{ec.europa.eu/eurostat}) and SDW is the statistical data warehouse by the European Central Bank (ECB, \href{https://sdw.ecb.europa.eu}{sdw.ecb.europa.eu}). The abbreviation \textit{s.a.} is short for seasonally adjusted.}
\end{tablenotes}
\end{threeparttable}
\end{small}
\end{center}
\end{table*}

\renewcommand\theequation{B.\arabic{equation}}
\section{Estimation of marginal distributions}
\begin{table*}[h]
\caption{Model selection criteria for fitting marginal distributions.}
\label{tab:bicdic}
\resizebox{\columnwidth}{!}{%
\begin{threeparttable}
\begin{tabular}{l|cc|cc|cc|cc}
\toprule
\multirow{2}{4em}{Country/ model} & \multicolumn{2}{c}{AT} & \multicolumn{2}{c}{BE} & \multicolumn{2}{c}{DE} & \multicolumn{2}{c}{ES} \\
  \cmidrule{2-9}
 & DIC & BIC & DIC & BIC & DIC & BI2 & DIC & BIC \\ 
  \hline
  Singh Maddala & \textbf{68596.84} & \textbf{68620.85} & 52888.25 & 52911.3 & 108715.76 & 108740.82 & 144238.69 & 144264.47 \\ 
  Dagum         & 68614.09 & 68638.12  & \textbf{52884.48} & \textbf{52907.63} & \textbf{108707.29} & \textbf{108732.43} & \textbf{143992.94} & \textbf{144019.12} \\ 
  Dagum 3       & \textbf{6464.7}& \textbf{76504.53} & \textbf{61404.13} & \textbf{61442.65} & \textbf{120871.22} & \textbf{120913.11} & \textbf{181203.14} & \textbf{181246.33}\\ 
  Log Normal 3  & 82605.86 & 82629.87 & 63796.69 & 63819.81 & 129146.5 & 129171.75 & 205399.91 & 205426.65 \\ 
  \bottomrule
\multirow{2}{4em}{Country/ model} & \multicolumn{2}{c}{FR}\ & \multicolumn{2}{c}{GR} & \multicolumn{2}{c}{IE} & \multicolumn{2}{c}{IT}  \\
  \cmidrule{2-9}
 & DIC & BIC & DIC & BIC & DIC & BIC & DIC & BIC \\ 
  \hline
  Singh Maddala & 295905.21 & 295934.24 & 65851.74 & 65876.27 & 129121.39 & 129147.17 & 185972.66 & 185999.65\\ 
  Dagum & \textbf{295019.32} & \textbf{295048.59} & \textbf{65476.29} & \textbf{65500.32} & \textbf{129052.7} & \textbf{129078.5} & \textbf{185410.91} & \textbf{185437.89} \\ 
  Dagum 3 & \textbf{339682.24} & \textbf{339728.59} & \textbf{71856.37} & \textbf{71896.41} & \textbf{140629.4} & \textbf{140672.24} & \textbf{213495.7} & \textbf{213540.73}\\ 
  Log Normal 3 & 364731.24 & 364759 & 76903.89 & 76927.91 & 155560.11 & 155585.88 & 223598.38 & 223625.4\\ 
 \bottomrule
\end{tabular}
\begin{tablenotes}[para,flushleft]
\scriptsize{\textit{Notes}: DIC stands for deviance information criterion and BIC is the abbreviation for Bayesian information criterion. Models with lower DIC or BIC are generally preferred, bold numbers indicate the lower value of the respective criterion.}
\end{tablenotes}
\end{threeparttable}}
\end{table*}

\clearpage
\renewcommand\theequation{C.\arabic{equation}}
\section{Additional macroeconomic VAR results}

\begin{figure}[!h]
     \centering
     \begin{subfigure}[b]{.88\textwidth}
        \caption{Standard monetary policy shock.}
        \label{fig:var_tg_app}
         \includegraphics[width=\textwidth, trim=10 0 10 10, clip]{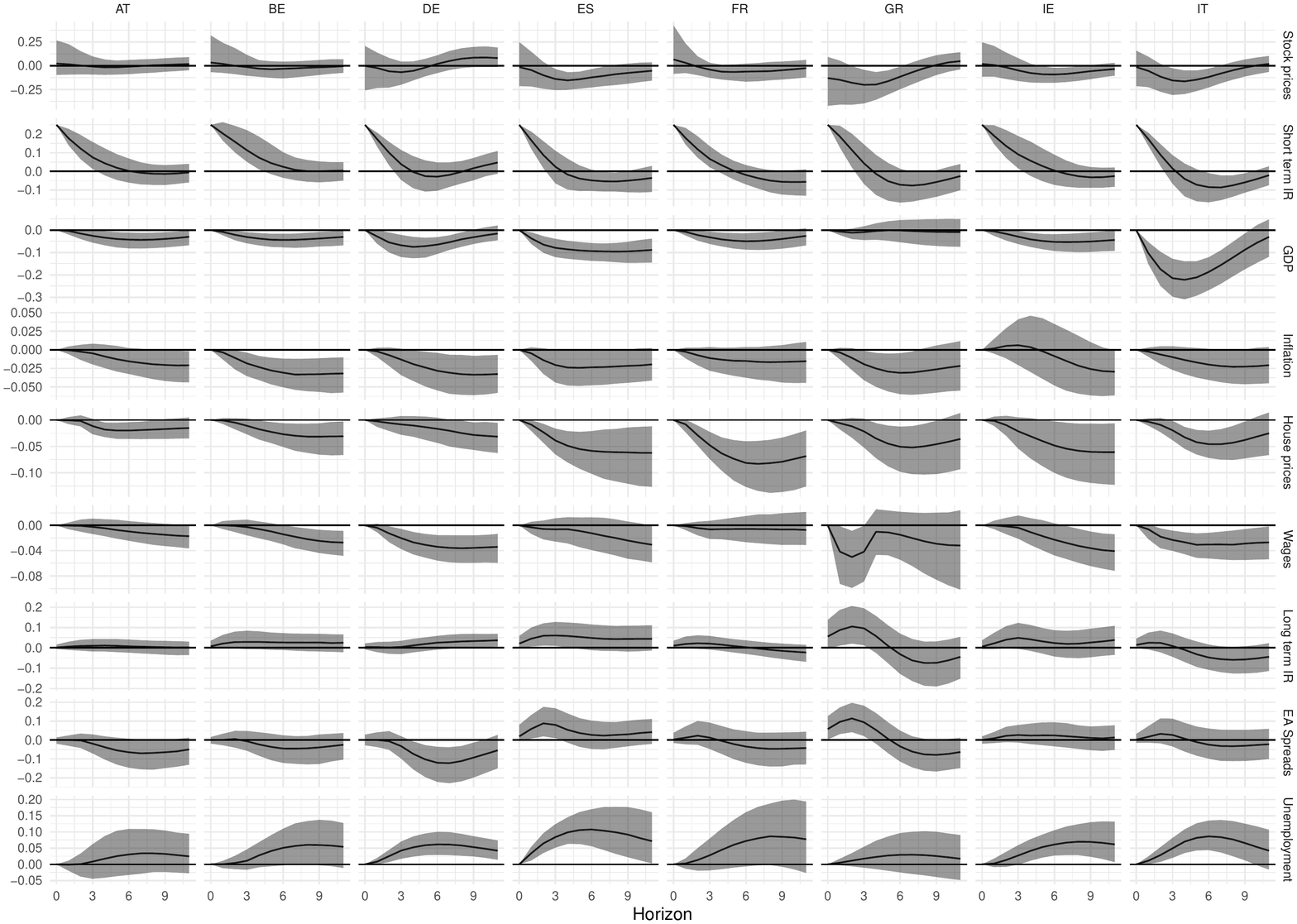}
     \end{subfigure}
     \begin{subfigure}[b]{.88\textwidth}
        \caption{Quantitative easing shock.}
        \label{fig:var_qe}
         \includegraphics[width=\textwidth, trim=10 0 10 10, clip]{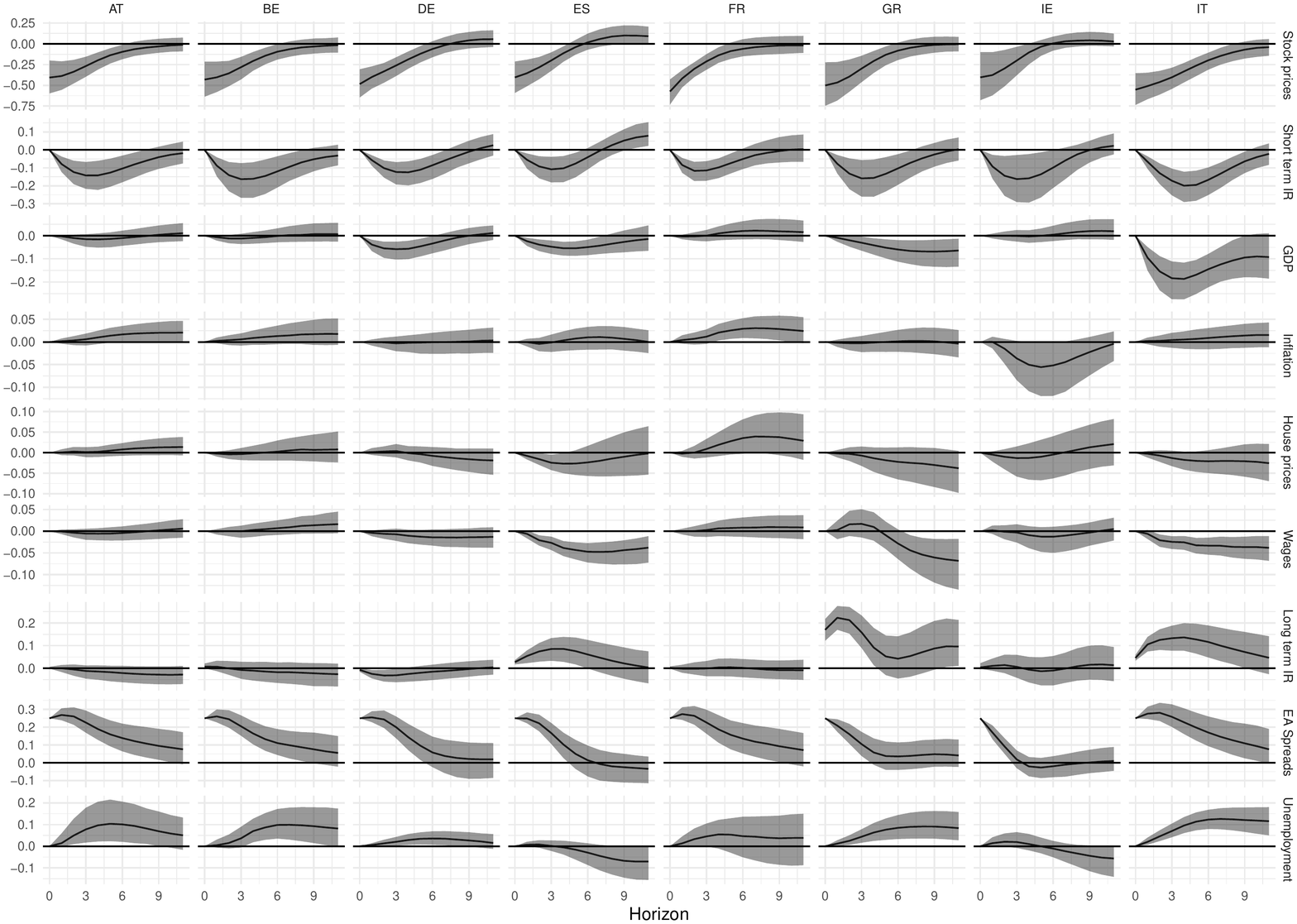}
     \end{subfigure}
        \caption{Impulse response functions from a BVAR to a contractionary one standard deviation standard monetary policy shock and quantitative easing shock.}
        \caption*{\footnotesize\textit{Notes}: Median response alongside the 68 percent posterior credible set. The black horizontal line marks zero.}
        \label{fig:varIRFs_app}
\end{figure}

\renewcommand\theequation{D.\arabic{equation}}
\section{Descriptives HFCS data}

\begin{figure}[!h]
     \centering
     \begin{subfigure}[b]{.88\textwidth}
        \caption{Income components per age groups.}
        \label{fig:inc_page}
         \includegraphics[width=\textwidth]{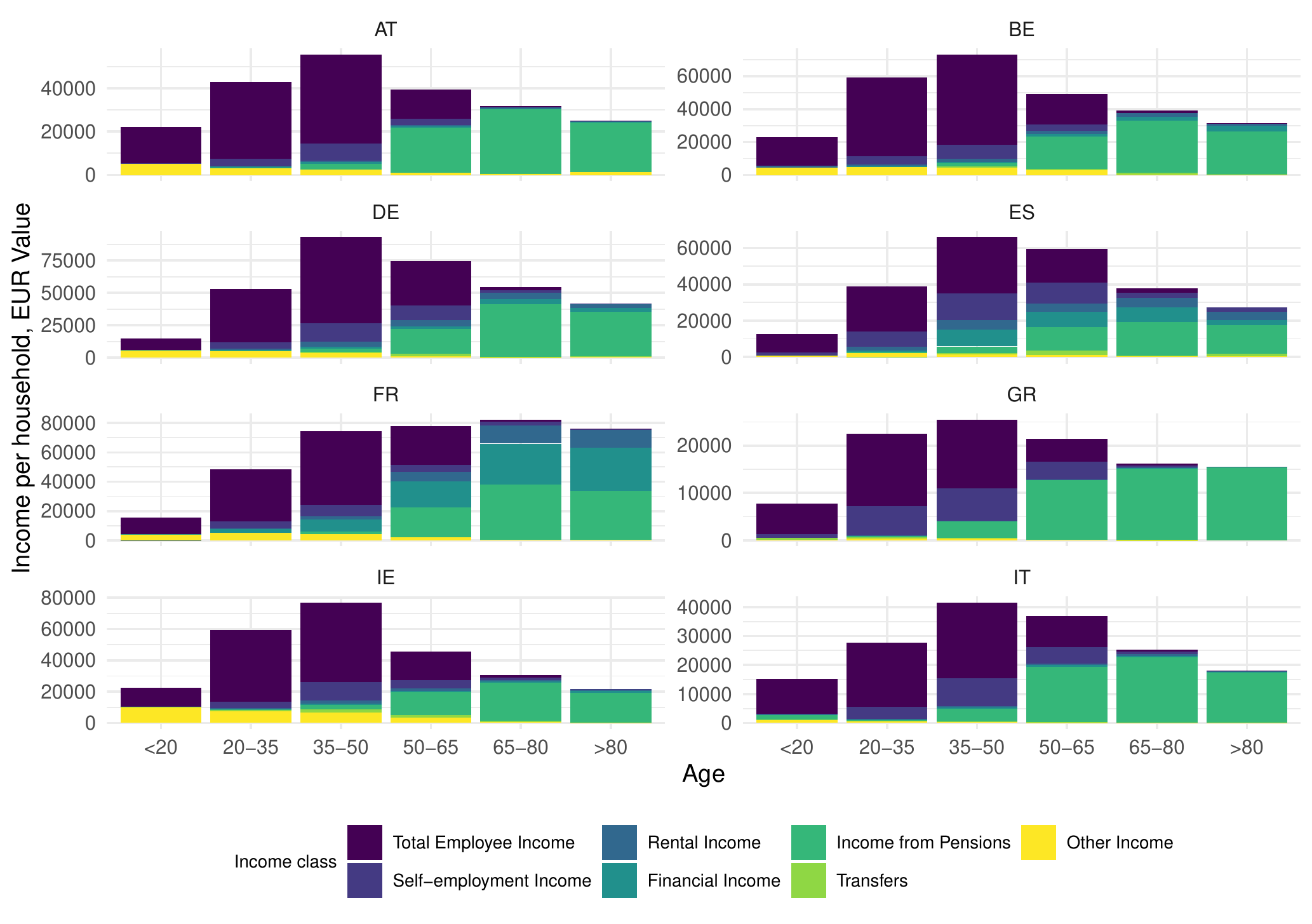}
     \end{subfigure}
     \begin{subfigure}[b]{.88\textwidth}
        \caption{Net wealth assets per age groups.}
        \label{fig:nw_page}
         \includegraphics[width=\textwidth]{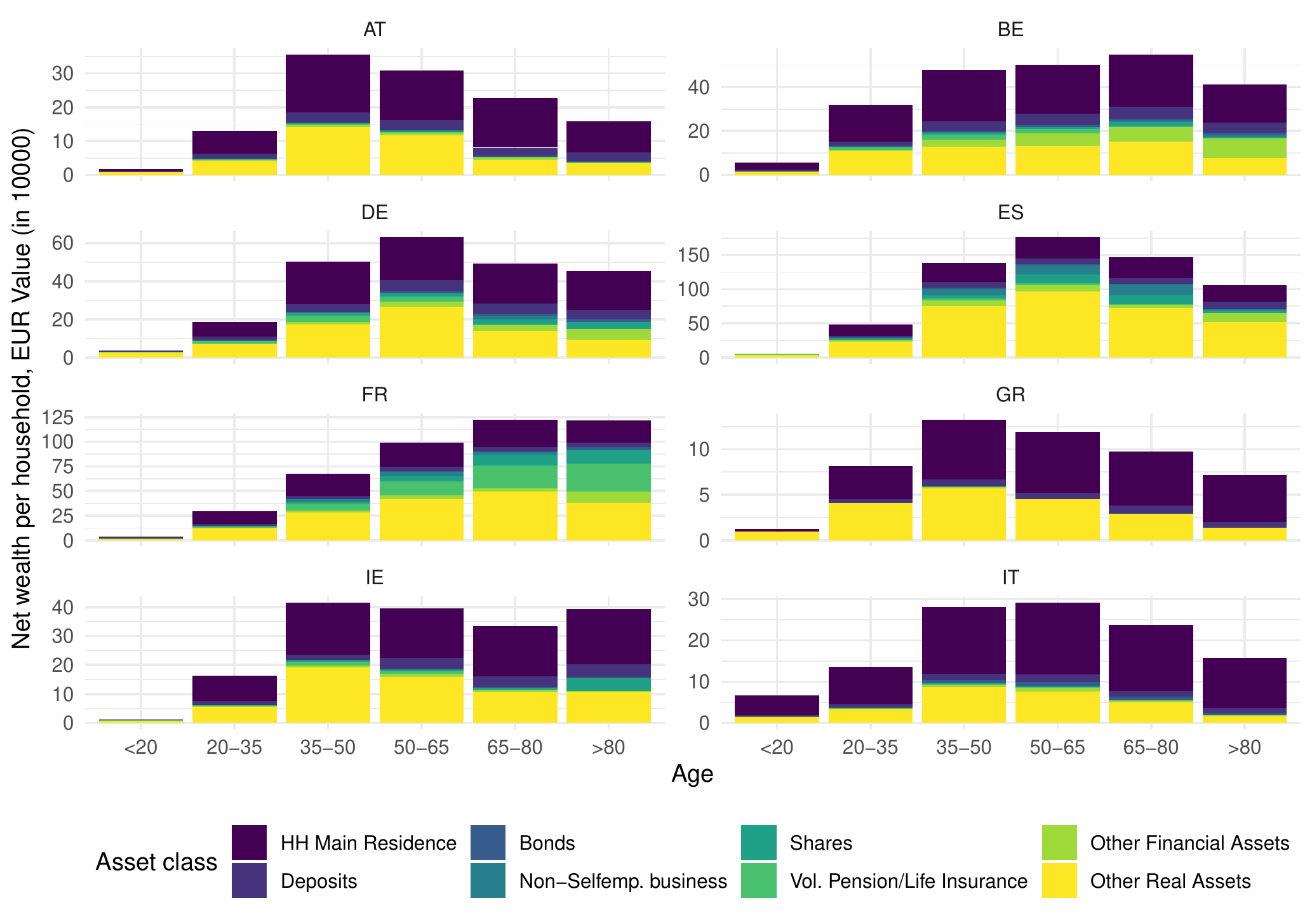}
     \end{subfigure}
        \caption{Distribution of types of income across age groups.}
        \caption*{\footnotesize\textit{Notes}: "HH main residence" denotes the household's main residence, "Non-Selfemp. business" is business from sources other than self-employment, "Vol. Pension/Life Insurance" denotes voluntary pensions or life insurances.}
        \label{fig:agegroups}
\end{figure}

\end{appendices}
\end{document}